\documentclass[prd,twocolumn,superscriptaddress,preprintnumbers,eqsecnum,showpacs,nofootinbib,
nobibnotes,noeprint]{revtex4-1}
\usepackage[latin1]{inputenc}
\usepackage{graphicx}
\usepackage{lipsum}
\usepackage{wrapfig, blindtext}
\usepackage{comment}
\usepackage{tabularx}
\usepackage{float}
\usepackage{amsfonts,bm}
\usepackage{paralist}
\usepackage[english]{babel}
\usepackage{slashed}
\usepackage[usenames]{xcolor}
\usepackage{mathtools}
\usepackage[makeroom]{cancel}
\usepackage{color}
\usepackage{slashed}
\usepackage{float}
\usepackage[caption=false, justification=centerlast]{subfig}
\usepackage[linkcolor=blue,citecolor=blue,urlcolor=blue,colorlinks=true]{hyperref} 
 \hypersetup{
           breaklinks=true,   
           colorlinks=true,   
           pdfusetitle=true,  
        }

\newcommand{\be}{\begin{equation}}
\newcommand{\bea}{\begin{eqnarray}}
\newcommand{\ee}{\end{equation}}
\newcommand{\eea}{\end{eqnarray}}

\def\1eq#1{Eq.~(\ref{#1})}

\def\2eqs#1#2{Eqs.~(\ref{#1}) and~(\ref{#2})}
\def\3eqs#1#2#3{Eqs.~(\ref{#1}),~(\ref{#2}) and~(\ref{#3})}

\newcommand{\Dr}{\mathcal Z} 

\def\ie{{\it i.e.}, }
\def\eg{{\it e.g.}, }
\def\s#1{{\scriptscriptstyle #1}}

\newcommand{\Lsym}{ \mathit{L}_{{sym}}}   
\newcommand{\Lasym}{ \mathit{L}_{{sg}}}

\def\s#1{{\scriptscriptstyle #1}}

\newcommand{\fatg}{{\rm{I}}\!\Gamma}

\newcommand{\Gnp}{\Gamma}

\begin{document}

\title{ Theory and phenomenology of  the three-gluon vertex}
\author{J. Papavassiliou}
\address{\mbox{Department of Theoretical Physics and IFIC, 
University of Valencia and CSIC},
E-46100, Valencia, Spain }
\author{ A.~C. Aguilar }
\address{ University of Campinas - UNICAMP, Institute of Physics ``Gleb Wataghin'', 
13083-859 Campinas, S\~{a}o Paulo, Brazil }
\author{  M.~N. Ferreira}
\address{ University of Campinas - UNICAMP, Institute of Physics ``Gleb Wataghin'', 
13083-859 Campinas, S\~{a}o Paulo, Brazil }

\begin{abstract}

  The three-gluon vertex is a fundamental ingredient of the intricate QCD dynamics,
  being inextricably connected to key 
  nonperturbative phenomena, such as the emergence of a mass scale in the gauge sector of
the theory. In this presentation we review the main theoretical properties of the
three-gluon vertex in the Landau gauge, obtained from the fruitful synergy between
functional methods and lattice simulations. We pay particular attention to 
the manifestation and origin of the infrared suppression of its main form factors and the
associated zero crossing. In addition, we discuss certain 
characteristic phenomenological applications that require this special vertex as input.

\end{abstract}


\maketitle

\section{Introduction}

The three-gluon vertex plays a pivotal role in the structure and dynamics of
Yang-Mills theories, reflecting their non-Abelian nature in the
form of the gauge boson self-interaction that it induces~\mbox{\cite{Marciano:1977su,Ball:1980ax,Davydychev:1996pb}}. In fact, the
most preeminent perturbative property of these theories, namely 
asymptotic freedom~\cite{Gross:1973id,Politzer:1973fx}, is intimately
linked to the action of this vertex.

In recent years, the QCD community has been gradually
unveiling the rich infrared facets of this vertex, which are  
instrumental to a wide array of nonperturbative phenomena;
for a representative set of references, see~\cite{Alkofer:2004it,Huber:2012zj,Pelaez:2013cpa,Aguilar:2013vaa,Blum:2014gna,Eichmann:2014xya,Mitter:2014wpa,Williams:2015cvx,Blum:2015lsa,Cyrol:2016tym,Corell:2018yil,Boucaud:2017obn,Huber:2018ned, Aguilar:2019jsj,Aguilar:2019uob,Aguilar:2019kxz,Parrinello:1994wd,Alles:1996ka,Parrinello:1997wm,Boucaud:1998bq,Cucchieri:2006tf,Cucchieri:2008qm,Athenodorou:2016oyh,Duarte:2016ieu,Boucaud:2017obn,Vujinovic:2018nqc}.
Several of these works have underscored 
the subtle interplay of the three-gluon vertex
with the two-point sector of the theory, and in particular
 the mass-generating patterns associated with the 
gluon and ghost propagators~\cite{Cornwall:1981zr, Alkofer:2008dt,Alkofer:2008jy, Aguilar:2008xm,Huber:2012zj,Pelaez:2013cpa,Aguilar:2013vaa,Blum:2014gna,Blum:2015lsa,Eichmann:2014xya,Vujinovic:2014fza,Mitter:2014wpa,Williams:2015cvx,Cyrol:2016tym}.
As a result, the  three-gluon vertex provides
an outstanding testing ground for a variety of physical ideas and
field-theoretic mechanisms~\cite{Aguilar:2021uwa,Eichmann:2021zuv,  Eichmann:2021miy,Meyers:2012ka, Souza:2019ylx,Huber:2021yfy}.
In this presentation we provide a synopsis of some of the
most important findings of this exploration.

The outline of this contribution is as follows. In Sec.~\ref{sec:genpr} we introduce the notation
and comment on the general
properties of the three-gluon vertex, give one of its standard tensorial decompositions, 
and report the Slavnov-Taylor identity (STI) that is satisfies.
In Sec.~\ref{sec:nonper} we discuss the
three main nonperturbative approaches used in the scrutiny of the three-gluon vertex, namely
functional methods, lattice simulations and STI-based constructions. 
Next, in Sec.~\ref{sec:supp} we analyse in some detail one of the most
exceptional nonperturbative features of the three-gluon vertex, namely the  
suppression of its predominant form factors for  Euclidean momenta 
comparable to the fundamental QCD scale, and the associated logarithmic infrared
divergence at the origin. In Sec.~\ref{sec:pheno} we discuss two phenomenological applications 
of the three-gluon vertex, namely ({\it a}) the effective charge obtained from it,
and ({\it b}) its impact on the computation of the mass of the pseudoscalar glueball. 
Finally, in Sec.~\ref{sec:conc} we summarize our conclusions.

\section{\label{sec:genpr}General properties}

We work in the Landau gauge,   where  the gluon propagator,  \mbox{$ \Delta^{ab}_{\mu\nu}\left(q\right) = -i\delta^{ab}\Delta_{\mu\nu}(q)$},  is fully transverse,  \ie
\be
\label{eq:prop}
\Delta_{\mu\nu}\left(q\right)  =  
 P_{\mu \nu}(q) \Delta(q^2)\,,  \quad \Delta(q^2) =  \Dr(q^2)/{q^2}\,,
\ee
where  $P_{\mu\nu}(q):=g_{\mu\nu}-{q_\mu q_\nu}/{q^2}\ $ is the usual transverse projector, and $\Delta(q^2)$ the scalar component of the gluon propagator. In addition, we have
defined the gluon dressing function, denoted by $\Dr(q^2)$.  

It is also convenient to  introduce the ghost propagator, \mbox{ $D^{ab}(q^2)=i\delta^{ab}D(q^2)$},  related to its dressing function, $F(q^2)$,  by 
\be
D(q^2) =F(q^2)/{q^2}\,.
\label{eq:ghost_dressing}
\ee

\begin{figure}[h]
\includegraphics[width=0.4\textwidth]{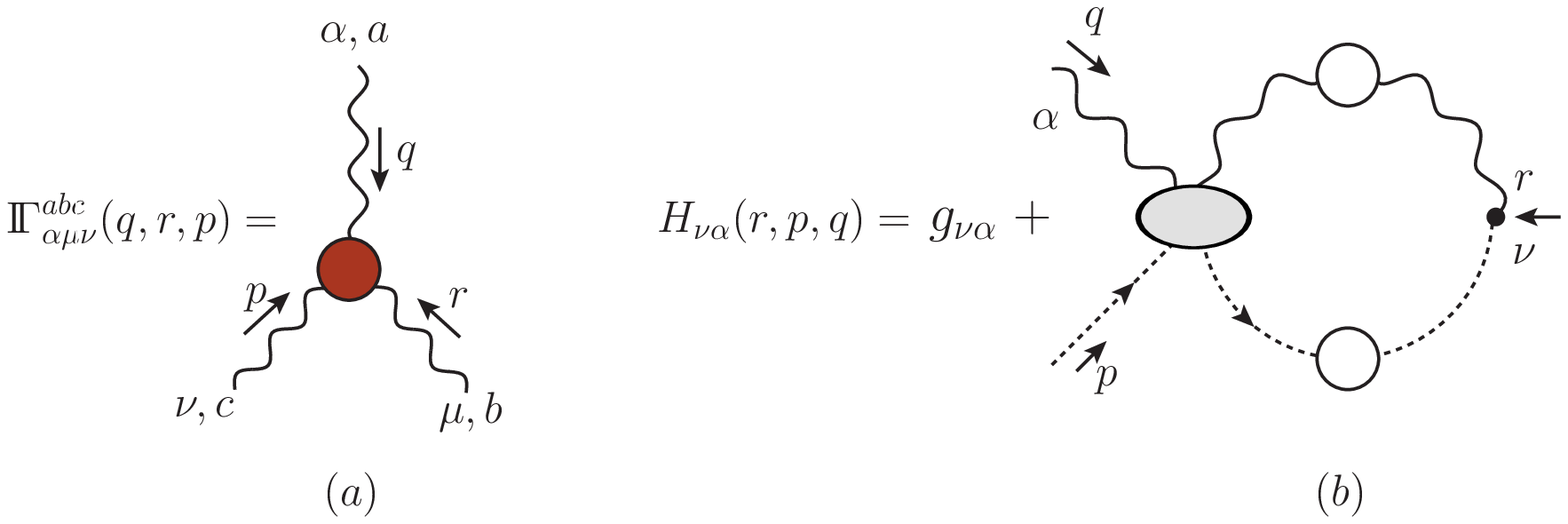}
\caption{The diagrammatic representations of the three-gluon vertex, $\fatg_{\alpha \mu \nu}^{abc}(q,r,p)$,
and the ghost-gluon scattering kernel,  $H_{\nu\alpha}(q,r,p)$, with the respective conventions of  momenta and indices.}
\label{fig:3gvertex}
\end{figure}

The full three-gluon vertex will be 
denoted by  \mbox{$\fatg^{abc}_{\alpha\mu\nu}(q,r,p)  = g f^{abc} \fatg^{\alpha\mu\nu}(q,r,p)$},  and is 
represented in Fig.~\ref{fig:3gvertex}, with \mbox{$q + p + r = 0$},  and  $g$  the gauge coupling. 

It is convenient to decompose $\Gnp_{\alpha\mu\nu}(q,r,p)$ into two distinct pieces~\mbox{~\cite{Ball:1980ax,Davydychev:1996pb,Aguilar:2019kxz}},  
\be
\fatg^{\alpha\mu\nu}(q,r,p) = \Gamma_{\!\s L}^{\alpha\mu\nu}(q,r,p) + \Gamma_{\!\s T}^{\alpha\mu\nu}(q,r,p)\,,
\label{decomp}
\ee
where  $\Gamma_{\!\s L}^{\alpha\mu\nu}(q,r,p)$ and $\Gamma_{\!\s T}^{\alpha\mu\nu}(q,r,p)$
are the  ``longitudinal'' and ``transverse'' parts of the three-gluon vertex,  respectively.   While the former
saturates the corresponding STIs [see \1eq{stip}],   the  latter is automatically
conserved when contracted by $q^\alpha$, $ r^\mu$,  and $p^\nu$, \ie 
 \mbox{ $q^\alpha \Gamma_{\!\s T}^{\alpha\mu\nu} = r^\mu  \Gamma_{\!\s T}^{\alpha\mu\nu} = p^\nu \Gamma_{\!\s T}^{\alpha\mu\nu}=0$}.
 
The tensorial decompositions of $\Gamma_{\!\s L}^{\alpha\mu\nu}(q,r,p)$ and $\Gamma_{\!\s T}^{\alpha\mu\nu}(q,r,p)$
reads 
\bea
\Gamma_{\!\s L}^{\alpha\mu\nu}(q,r,p) &=& \sum_{i=1}^{10} X_i(q,r,p) \ell_i^{\alpha\mu\nu} \,, \nonumber \\ 
 \Gamma_{\!\s T}^{\alpha\mu\nu}(q,r,p) &=& \sum_{i=1}^{4}Y_i(q,r,p)t_i^{\alpha\mu\nu} \,,
\label{BCbasis}
\eea
where the explicit expressions of the basis elements $\ell_i^{\alpha\mu\nu}$ and $t_i^{\alpha\mu\nu}$
are given in Eqs.~(3.4) and~(3.6) of~\cite{Aguilar:2019jsj}, respectively.

Another familiar quantity  introduced  in the studies of the three-gluon vertex is the {\it transversally projected vertex}, 
 $\overline{\Gamma}_{\alpha \mu \nu}(q,r,p)$, defined as~\cite{Eichmann:2014xya,Huber:2018ned} 
\be
\label{eq:Gammabar}
\!\!\overline{\fatg}_{\alpha \mu \nu}(q,r,p) \!=\!\fatg^{\alpha\!' \mu\!' \nu\!'}\!(q,r,p)P_{\alpha'\! \alpha}(q) P_{\mu'\! \mu}(r) P_{\nu'\! \nu}(p)  \,.
\ee

 In addition,  we define the tree-level counterpart of Eq.~\eqref{eq:Gammabar}, 
\be
\label{eq:Gammabar0}
\overline{\Gamma}_{\alpha \mu \nu}(q,r,p) =\Gamma_{\!0}^{\alpha\!' \mu\!' \nu\!'}\!(q,r,p)P_{\alpha'\! \alpha}(q) P_{\mu'\! \mu}(r) P_{\nu'\! \nu}(p),  
\ee
where $\Gamma_{\!0}^{\alpha\!' \mu\!' \nu\!'}\!(q,r,p)$  is the  standard tree-level expression,   given by 
\be
\Gamma_{\!0}^{\alpha\mu\nu}= (q-r)^{\nu}g^{\alpha\mu} + (r-p)^{\alpha}g^{\mu\nu} + (p-q)^{\mu}g^{\alpha\nu}\,;
\label{treelevel}
\ee
it may be obtained from Eq.\,(\ref{decomp}) by setting \mbox{$X_1=X_4=X_7 = 1$}, and zero for all other
form-factors. 

The STI satisfied by $\fatg_{\alpha\mu\nu}(q,r,p)$ reads
\be
p^\nu \fatg_{\alpha \mu \nu}(q,r,p) = F(p^2) [ {\cal T}_{\mu\alpha}(r,p,q) - {\cal T}_{\alpha\mu}(q,p,r) ] \,,
\label{stip}
\ee
with 
\be
{\cal T}_{\mu\alpha}(r,p,q) := \Delta^{-1}(r^2) P_\mu^\sigma(r) H_{\sigma\alpha}(r,p,q)\,,
\label{defAT}
\ee
where $H_{\nu\mu}(q,p,q)$ denotes the ghost-gluon scattering kernel, represented diagrammatically
in the panel ({\it b}) of Fig.~\ref{fig:3gvertex}. Its 
tensorial decomposition is given by~\cite{Ball:1980ax,Davydychev:1996pb,Aguilar:2018csq} 
\bea 
H_{\nu\mu}(q,p,r) &=& g_{\nu\mu} A_1 + q_\mu q_\nu A_2 + r_\mu r_\nu A_3  \nonumber  \\  
&+&q_\mu r_\nu A_4 + r_\mu q_\nu A_5 \,,
\label{theAi}
\eea
where we use the compact notation \mbox{$A_i:= A_i(q,p,r)$}.

\section{\label{sec:nonper} Nonperturbative methods}

The rich kinematic structure of the three-gluon vertex 
makes its nonperturbative study particularly challenging.
There are three main frameworks for dealing with this problem: ({\it i})
 Functional methods,  such as the Schwinger-Dyson equations (SDEs)~\mbox{\cite{Schleifenbaum:2004id,Huber:2012kd,Aguilar:2013xqa,Huber:2012zj,Blum:2014gna,Eichmann:2014xya,Williams:2015cvx, Binosi:2016wcx}} and the
functional renormalization group~\mbox{\cite{Corell:2018yil,Cyrol:2017ewj,Cyrol:2016tym}}; ({\it ii}) large-volume lattice simulations~\mbox{\cite{Parrinello:1997wm,Boucaud:1998bq,Cucchieri:2006tf,Cucchieri:2008qm,Athenodorou:2016oyh,Duarte:2016ieu,Boucaud:2017obn,Vujinovic:2018nqc,Aguilar:2019uob,Aguilar:2021lke}};
and ({\it iii}) STI-based 
reconstructions of the longitudinal part, $\Gamma_{\!\s L}^{\alpha\mu\nu}(q,r,p)$,
in the spirit of the ``gauge-technique''~\mbox{\cite{Salam:1963sa,Salam:1964zk,Delbourgo:1977jc,Delbourgo:1977hq}}.

\begin{figure}[h]
\includegraphics[scale=0.35]{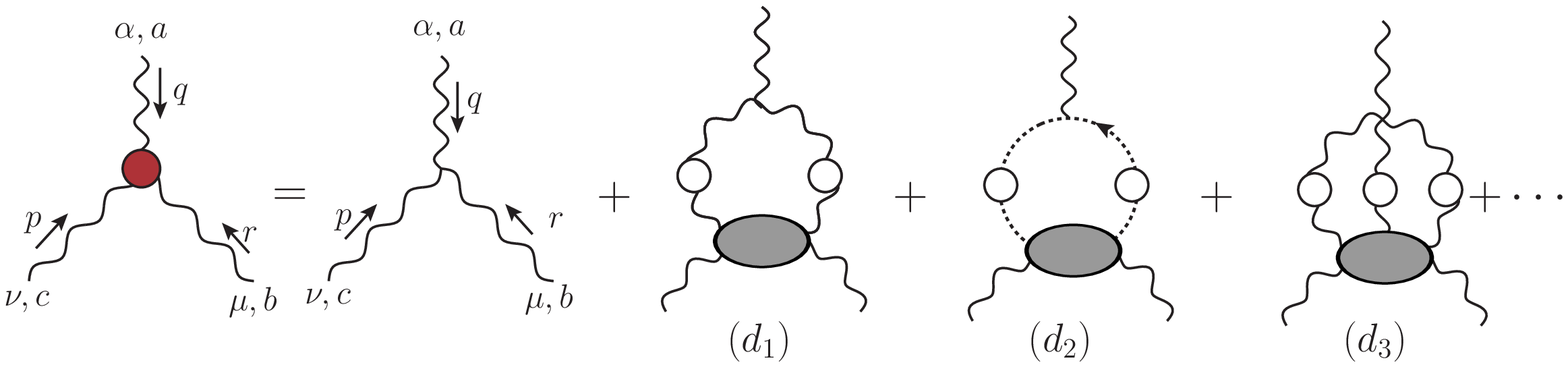}
\vspace{-0.25cm}
\caption{ The SDE of the three-gluon vertex. The white (gray) circles (ellipses) indicate fully dressed propagators (kernels), while the dots indicate the omitted terms.}
\label{fig:3g_SDE_kernel}
\end{figure}

({\it i}) \emph{ \underline{Functional methods}}: The diagrammatic representation of the SDE that governs the evolution of the three-gluon vertex 
is shown in Fig.~\ref{fig:3g_SDE_kernel}. The self-consistent treatment of this equation is particularly complicated,
and entails its coupling to additional related equations, such as the SDEs of the gluon and ghost propagators.
In practice, this task is considerably simplified by using as inputs the lattice results for 
$\Delta(q^2)$ and $F(q^2)$.

 ({\it ii})  \emph{ \underline {Lattice simulations}}: In this case the three-gluon vertex
   is accessed through the functional averaging of the quantity 
   $\langle \widetilde{A}^a_\alpha(q) \widetilde{A}^b_\mu(r) \widetilde{A}^c_\nu(p) \rangle$,
   where $\widetilde{A}^a_\alpha(q)$ denotes the SU(3) gauge field. Specifically,
   the connected three-point function, ${\cal G}_{\alpha \mu \nu}(q,r,p)$, defined as 
\be
\label{eq:gGD3}
{\cal G}_{\alpha \mu \nu}(q,r,p) = g \overline{\fatg}_{\alpha \mu \nu}(q,r,p) \Delta(q^2) \Delta(r^2) \Delta(p^2) \ ,
\ee
 is given by \mbox{
$\langle \widetilde{A}^a_\alpha(q) \widetilde{A}^b_\mu(r) \widetilde{A}^c_\nu(p) \rangle  =  f^{abc} {\cal G}_{\alpha \mu \nu}(q,r,p)$}.  $\overline{\fatg}_{\alpha \mu \nu}(q,r,p)$ is finally obtained 
after an appropriate amputation of the gluon propagators. 

The typical structure of lattice ``observables'' is
\be
L(q,p,r) =  \frac{W^{\alpha \mu \nu}(q,r,p)\overline{\fatg}_{\alpha \mu \nu}(q,r,p) }
{\rule[0cm]{0cm}{0.45cm}W_{\alpha \mu \nu}(q,r,p) W^{\alpha \mu \nu}(q,r,p)} 
\rule[0cm]{0cm}{0.5cm} \;,
\label{eq:latobs}
\ee
where the $W^{\alpha\mu\nu}(q,r,p)$ are appropriately chosen projectors~\cite{Athenodorou:2016oyh,Duarte:2016ieu,Boucaud:2017obn}.  
In what follows we will focus our attention on two special kinematic limits
involving a single momentum variable.

{\it (a)}  \emph{Soft  limit},  corresponding to the kinematic choice 
  \be
q\to 0 \,,\quad p=-r\,, \quad \theta:=\widehat{pr}=\pi \,,
\label{defasym}
\ee
obtained by setting \mbox{$W^{\alpha\mu\nu}(q,r,p) \to 2r^{\alpha} P^{\mu\nu}(r)$}, namely 
\be
L_{sg} (r^2) =  \frac{\Gamma_{0}^{\alpha\mu \nu}(q,r,p)\overline{\fatg}_{\alpha \mu \nu}(q,r,p)}
{\rule[0cm]{0cm}{0.45cm}\; {\Gamma_{0}^{\alpha\mu \nu}(q,r,p)\overline{\Gamma}_{\alpha \mu \nu}(q,r,p)}}
\rule[0cm]{0cm}{0.5cm} \Bigg|_{\substack{\!\!q\to 0 \\ p\to -r}}\,.
\label{Lsg}
\ee
  
{\it (b)} \emph{Totally symmetric} limit, 
\be
q^2 = p^2= r^2 :=s^2\,, \quad   \theta:= \widehat{q r} = \widehat{q p} = \widehat{r p} = 2\pi/3\,; 
\label{defsym}
\ee
the corresponding \mbox{$W^{\alpha\mu\nu}(q,r,p)$} and the expression for $\Lsym(s^2) $  may be found in Eqs. (2.18) and~(2.19)  of~\cite{Aguilar:2019uob}.

\vspace{0.5cm}

 ({\it iii})  \emph{\underline{STI}}:
As was first shown in~\cite{Ball:1980ax}, the STI of   
Eq.~\eqref{stip}, together with its cyclic permutation, determines the form factors $X_i(q,r,p)$ in terms
of the kinetic part of the gluon propagator, to be denoted by $J(q^2)$, the ghost dressing function, and three form factors of the
ghost-gluon kernel.

\begin{figure*}[t]
\begin{minipage}[b]{0.45\linewidth}
\centering
\hspace{-1.0cm}
\includegraphics[scale=0.9]{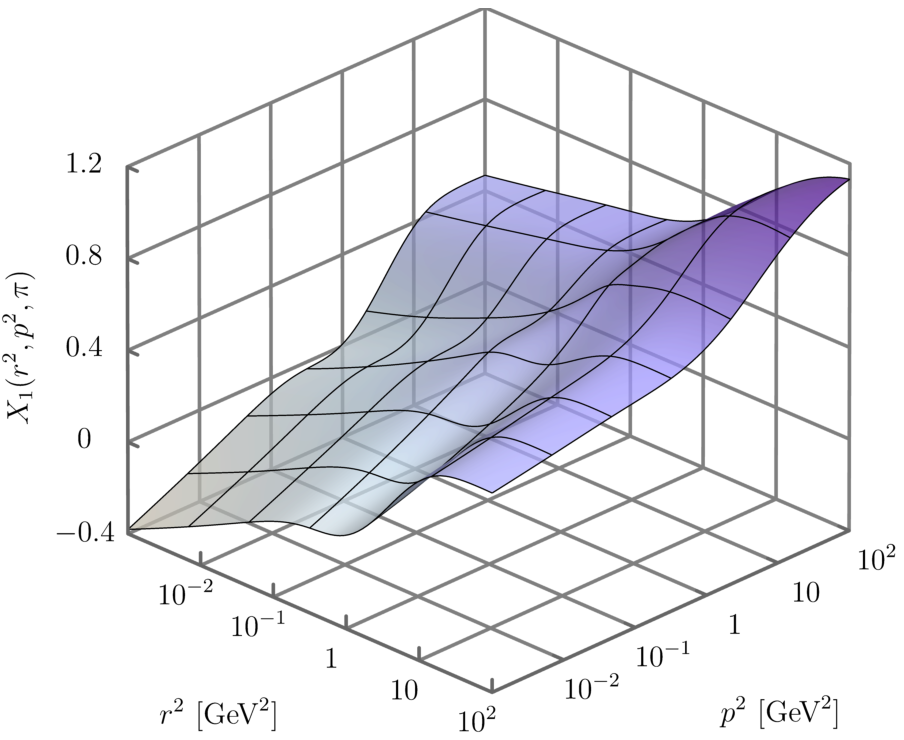}
\end{minipage}
\hspace{0.25cm}
\begin{minipage}[b]{0.45\linewidth}
\includegraphics[scale=0.9]{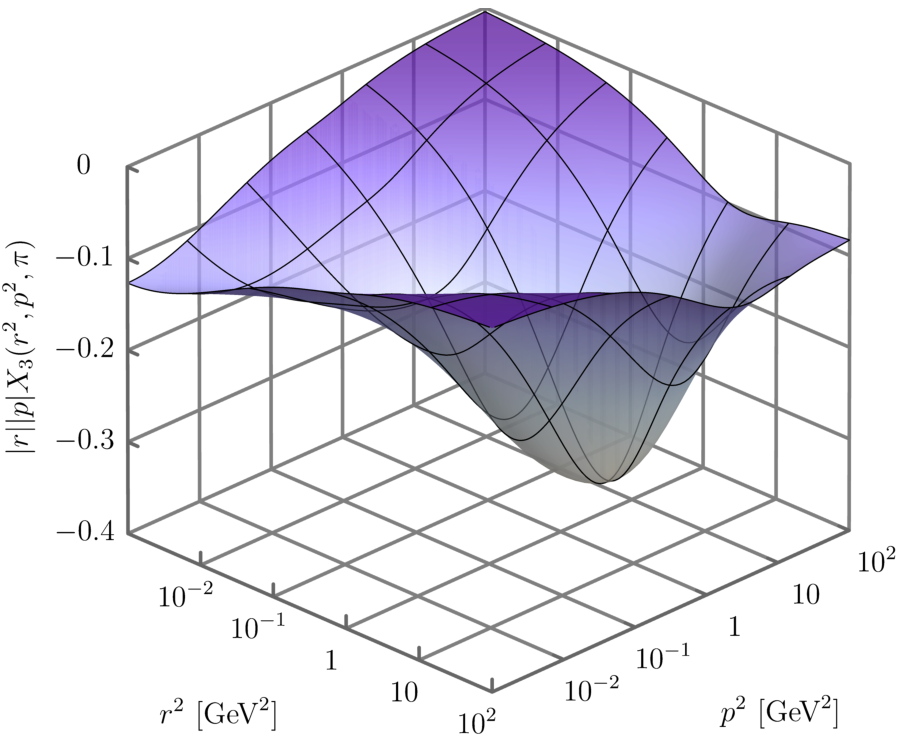}
\end{minipage}
\caption{A representative case of the three-gluon form factor  $X_1(r^2,p^2, \theta)$ (left panel) 
and for   $ |p||r|X_3(r^2,p^2, \theta)$  (right panel) for a fixed value of the angle,  \mbox{$\theta:=\widehat{pr}=\pi$}.  } 
\label{fig:3general}
\end{figure*}

Specifically, 
\begin{align}
&X_1 = \frac{1}{4}[(q^2-r^2)(b_{rpq} + b_{pqr} - b_{qpr} - b_{prq})]\,, \nonumber\\
&+ \!2(a_{pqr} \!+\! a_{prq}) \!+\! p^2(b_{qrp} \!+\! b_{rqp}) \!+\! 2(\,q\cdot p\,d_{prq} \!+\! \,r\cdot p\,d_{pqr}) ]\,, \nonumber\\
&X_2 = \frac{1}{4}[2(a_{prq} - a_{pqr})  -(q^2 - r^2)(b_{qrp} + b_{rqp}) \nonumber\\  
&+ 2(\,q\cdot p\,d_{prq} - \,r\cdot p\,d_{pqr}) + p^2(b_{prq} - b_{pqr} + b_{qpr} - b_{rpq})]\,,
\nonumber
\\
&X_3 = \frac{1}{q^2 - r^2}[a_{rpq} - a_{qpr} + r\cdot p \, d_{qpr} - q\cdot p \, d_{rpq}]\,, \nonumber
\\
&X_{10} = - \frac{1}{2}[b_{qrp} + b_{rpq} + b_{pqr} - b_{qpr} - b_{rqp} - b_{prq}],
\label{eq:X_sol}
\end{align} 
where we introduced the following compact notation 
\begin{align}
a_{qrp} := & F(r)J(p)A_1(p,r,q)\,,
\nonumber\\
b_{qrp} := & F(r)J(p)A_3(p,r,q)\,,
\nonumber\\
d_{qrp} := & F(r)J(p)[A_4(p,r,q)-A_3(p,r,q)]\,. 
\label{eq:ball_chiu_functions}
\end{align}
Due to the Bose symmetry of the three-gluon vertex,
the  remaining six $X_i$ may be computed by permuting the arguments appropriately (see Eq. (3.8) of~\cite{Aguilar:2019jsj}).

It is important to emphasize that in the original work of~\cite{Ball:1980ax} the kinetic term of the gluon propagator
was defined as $\Delta^{-1}(q^2) = q^2 J(q^2)$, while in the nonperturbative generalization presented in~\cite{Aguilar:2019jsj}
we have \mbox{$\Delta^{-1}(q^2) = q^2 J(q^2)+m^2(q^2)$}, where $m^2(q^2)$ is the running gluon mass~\cite{Cornwall:1981zr, Aguilar:2008xm, Binosi:2009qm,Cornwall:1981zr, Aguilar:2008xm,Roberts:2020hiw}.

Two representative results for the form factors $X_1(r^2, p^2, \theta$)  and  $|q||r|X_3(r^2, p^2, \theta$), obtained with Eq.~\eqref{eq:X_sol},   are shown on Fig.~\ref{fig:3general}.   In this figure  we present both form factors  as a function of the two momenta $p^2$ and $r^2$ when the angle  between these two momenta is fixed at the value $\theta= \pi$.   

Clearly, one can see that $X_1(r^2, p^2, \theta)$ has a completely nontrivial structure, which
persists for general values of the angle. Evidently, the most striking feature of this result is the
reduction of the size of $X_1(r^2, p^2, \theta)$ with respect to its tree-level value (unity); this effect 
is known in the literature
as ``infrared suppression''~\mbox{\cite{Aguilar:2013vaa,Athenodorou:2016oyh,Boucaud:2017obn,Blum:2015lsa,Corell:2018yil,Huber:2018ned, Aguilar:2019jsj}}.

Let us also point out that 
the projection of the three gluon vertex in the  totally symmetric 
limit,  defined in Eq.~\eqref{defsym},  can be written as~\cite{Aguilar:2019jsj}   
\bea  
\label{eq:GammaSym_Xi}
\Lsym (s^2)&=& X_1(s^2) - \frac{s^2}{2} X_3(s^2) \nonumber \\  &+& \frac{s^4}{4} Y_1(s^2) - \frac{s^2}{2} Y_4(s^2) \,.
\eea

\begin{figure*}[t]
\begin{minipage}[b]{0.45\linewidth}
\centering
\hspace{-1.0cm}
\includegraphics[scale=0.25]{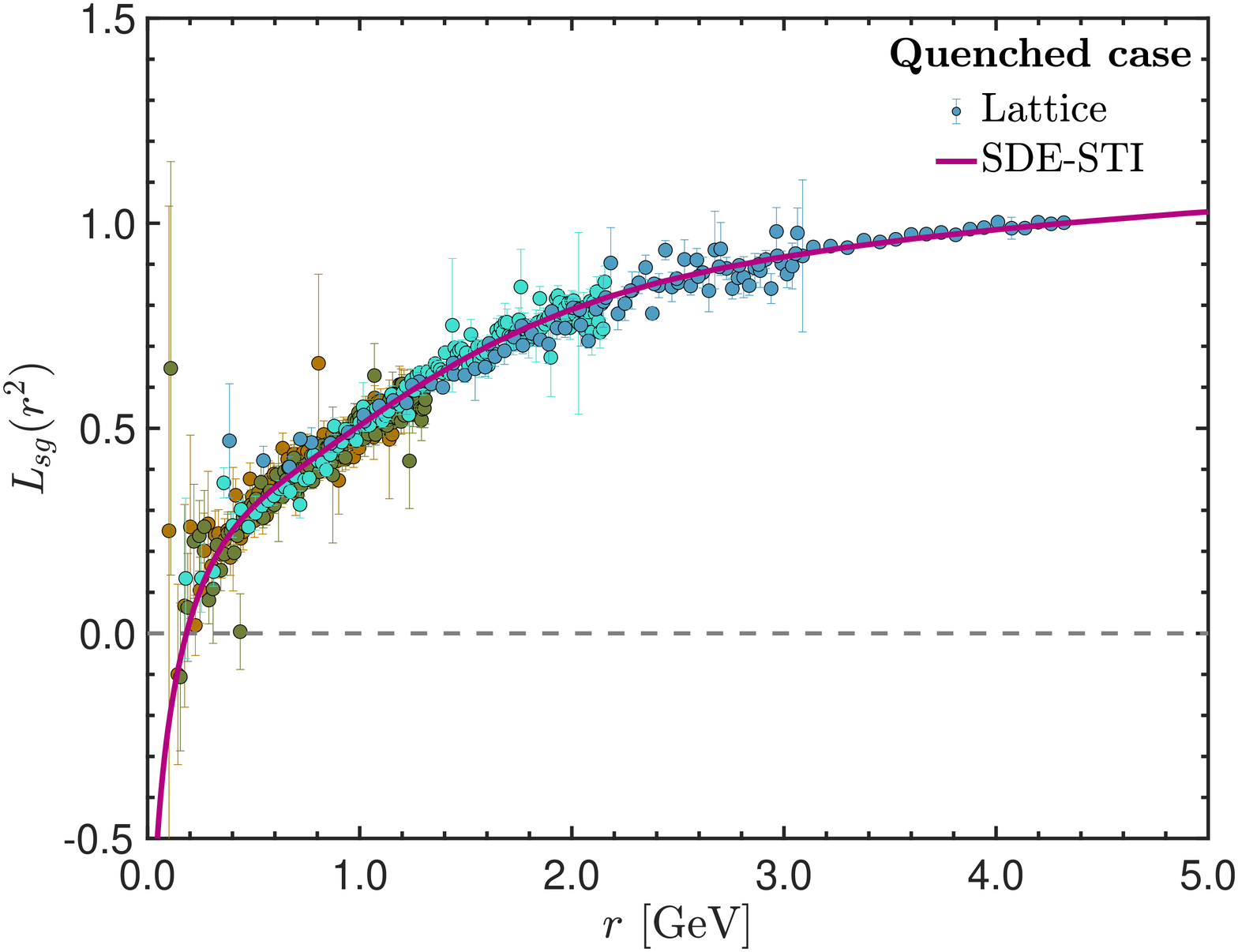}
\end{minipage}
\hspace{0.25cm}
\begin{minipage}[b]{0.45\linewidth}
\includegraphics[scale=0.25]{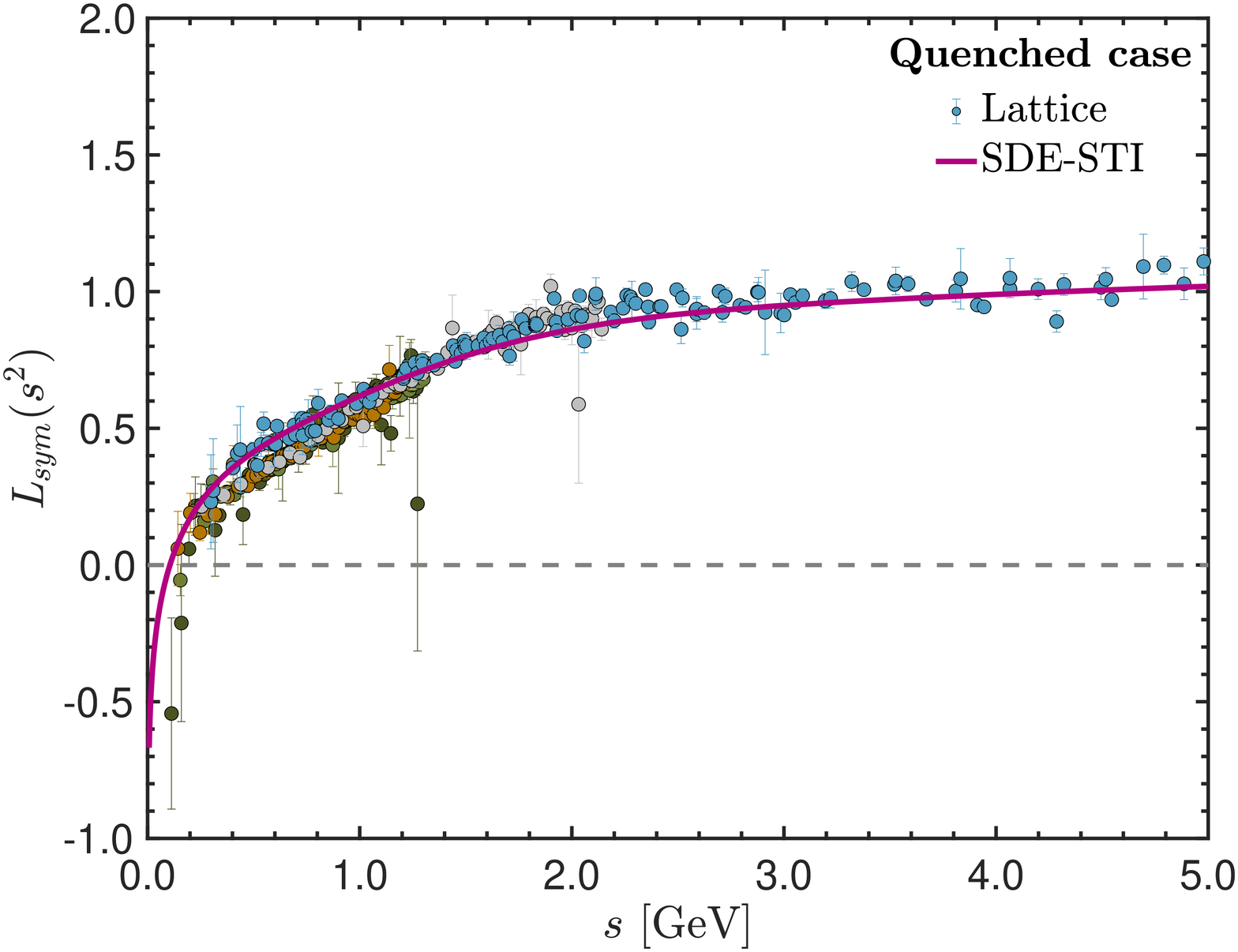}
\end{minipage}
\caption{ The projection of the three-gluon vertex in the  soft gluon,   $\Lasym(r^2)$  (left) and in  the   symmetric, $\Lsym(s^2)$,  (right) kinematic configurations  obtained  from lattice QCD of~\cite{Athenodorou:2016oyh,Boucaud:2017obn,Aguilar:2021lke} (solid circles) and  from the SDE-STIs approach  (magenta continuous curve)~\cite{Aguilar:2021lke}. } 
\label{fig:sym_asym}
\end{figure*}

\begin{figure*}[t]
\begin{minipage}[b]{0.45\linewidth}
\centering 
\hspace{-1.5cm}
\includegraphics[scale=0.25]{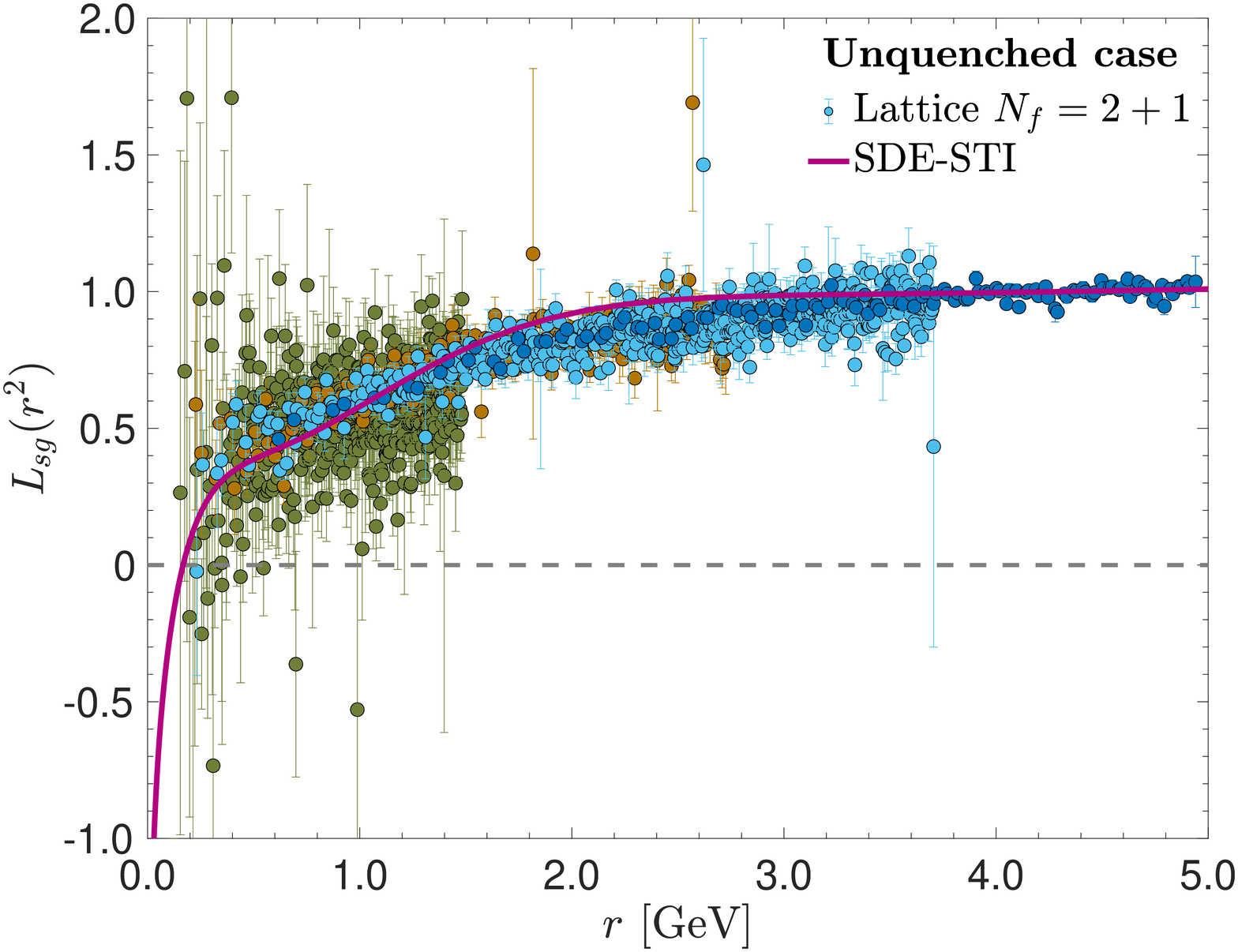}
\end{minipage}
\hspace{0.15cm}
\begin{minipage}[b]{0.45\linewidth}
\includegraphics[scale=0.25]{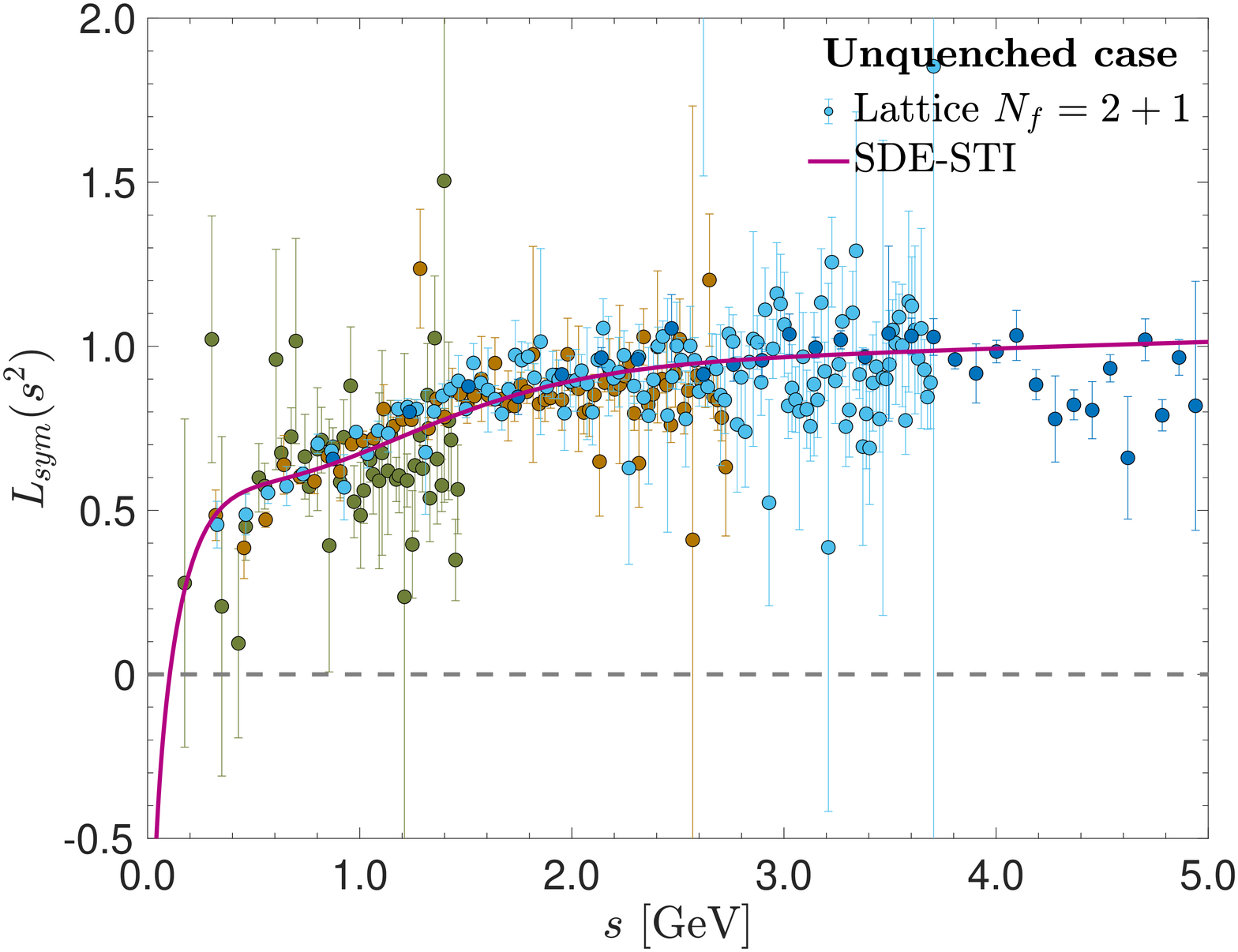}
\end{minipage}
\caption{ Left panel:  The unquenched  \mbox{$\Lasym(r^2)$} (left) and $\Lsym(s^2)$ (right) obtained from lattice QCD with  $N_f=2+1$ dynamical quarks~\cite{Aguilar:2019uob},  and  from the SDE-STI  approach (magenta  continuous curve).  }
\label{fig:unquenched}
\end{figure*}

On other hand, for the case of the soft limit configuration of \1eq{defasym}, the expression for 
$\Lasym (r^2)$ is given by 
\be
\label{eq:asyGamma}
 \Lasym (r^2)=  X_1(r^2,r^2, \pi) - r^2X_3(r^2,r^2, \pi)\,\,;
\ee
note that the result is free of transverse form factors $Y_i$.

Notice that the soft gluon kinematic limit of $X_1$ and $X_3$ corresponds to the curves that lie on the diagonal ``slice'' of the 3D plots of Fig.~\ref{fig:3general} where \mbox{$p^2=r^2$}.    In the left panel of  Fig.~\ref{fig:sym_asym} we show a comparison of the  $\Lasym (r^2)$  computed using the  SDE-STI  approach (magenta  continuous curve) and a combination of the lattice data of~\cite{Athenodorou:2016oyh,Boucaud:2017obn,Aguilar:2021lke} (solid circles), for the case of quenched QCD.
It is clear that both methods 
corroborate the infrared suppression of the three-gluon vertex.   In the right panel we show the results for  $ \Lsym (s^2)$,
obtained when we set $Y_i = 0$ in Eq.~\eqref{eq:GammaSym_Xi}. Once again the coincidence with the lattice data is rather notable, and the presence of the steep decline in the infrared is visible in both approaches. 
In addition, the same pattern (suppression and zero crossing) persists qualitatively unaltered
when $N_f=2+1$  dynamical quarks are added~\cite{Aguilar:2019uob}, as can be clearly seen in Fig.~\ref{fig:unquenched}.

\section{\label{sec:supp}Infrared suppression}

One of the most remarkable nonperturbative features of the three-gluon vertex in the Landau gauge is its 
infrared suppression, as established clearly in the results of the previous section. 
Thus, form factors such as $X_1$, $X_4$, and $X_7$, 
which, due to renormalization, acquire their tree level value (unity) at $4.3$ GeV, reduce their size to half at around \mbox{$1$ GeV}. 
This tendency culminates with a characteristic 
reversal of the sign, known as ``zero crossing''~\mbox{\cite{Aguilar:2013vaa,Eichmann:2014xya,Blum:2014gna,Vujinovic:2014fza,Athenodorou:2016oyh,Boucaud:2017obn}},  
followed by a logarithmic divergence of the corresponding form factor at the origin.

\begin{figure}[h]
\centering
\includegraphics[scale=0.25]{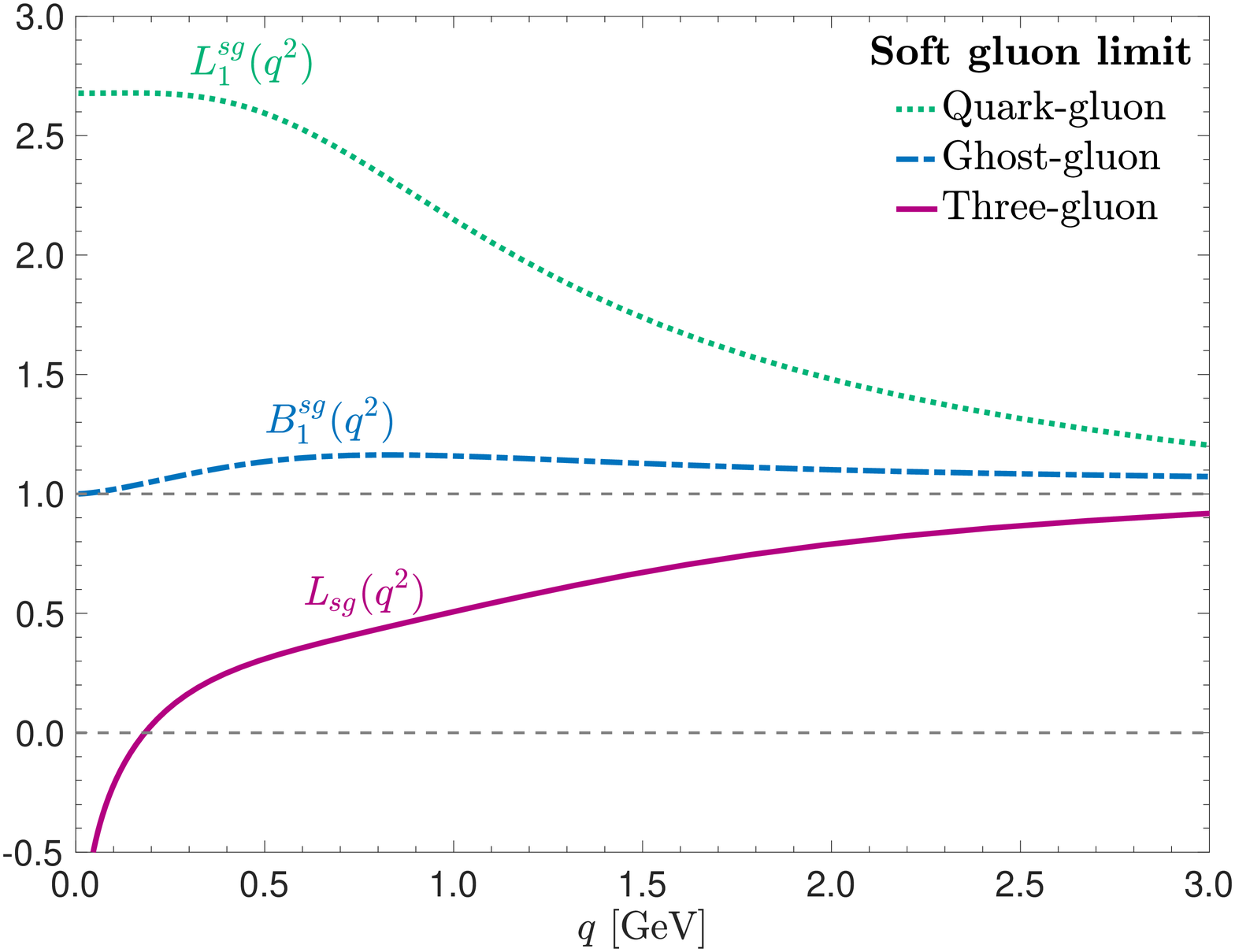}
\caption{  The behavior of the form factors,  $L_1^{sg}(q^2)$,  $B_1^{sg}(q^2)$,  and $\Lasym(q^2)$  associated with the  classical tensor structures of the 
quark-gluon (green dotted),  ghost-gluon (blue dashed) and three-gluon (magenta continuous) vertices,  respectively,  in the soft gluon limit. }
\label{fig:supvert}
\end{figure}

This type of behavior is in sharp contradistinction to what happens with the other vertices of the theory that
have been explored so far, such as the quark-gluon or the ghost-gluon vertex.
Indeed, as one can see in Fig.~\ref{fig:supvert}, the analogous form factors
display a clear enhancement for the same range of intermediate and infrared momenta.

\begin{figure}[h]
\centering
\includegraphics[scale=0.37]{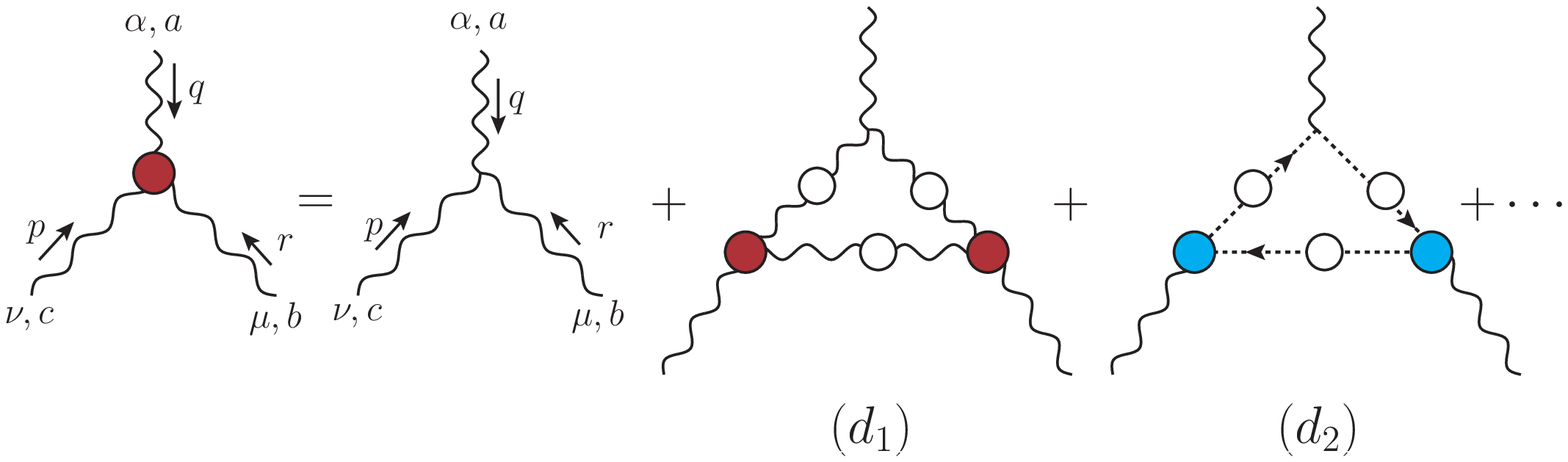}
\caption{ The SDE of the three-gluon vertex at the one-loop dressed level. The white (red and blue) circles indicate fully dressed propagators (vertices). The diagrams $(d_1)$ and $(d_2)$ are the 
gluon and the ghost triangle
contributions  entering in the skeleton expansion of three-gluon vertex. }
\label{fig:supression}
\end{figure}

From the theoretical
point of view, this particular feature of the three-gluon vertex 
hinges on the subtle interplay between dynamical effects 
originating from the two-point sector of the theory~\cite{Roberts:1994dr,Alkofer:2000wg,Fischer:2006ub,Cloet:2013jya,Aguilar:2015bud,Binosi:2014aea}. This may be understood at the level of the one-loop dressed version of the SDE in Fig. ~\ref{fig:3g_SDE_kernel}, which is shown in Fig. ~\ref{fig:supression}.
The crucial theoretical ingredient is that, whereas the gluon acquires dynamically an effective mass,
the ghost remains massless even nonperturbatively. 
As a result, the loops of the three-gluon vertex 
containing gluons (such as the ($d_1$) in Fig. ~\ref{fig:supression}) give rise to ``protected'' logarithms, because the effective gluon mass $m$
acts as an infrared regulator. 
Instead, loops containing ghosts (such as the ($d_2$) in Fig.~\ref{fig:supression})
produce ``unprotected'' logarithms, which diverge at the origin~\cite{Aguilar:2013vaa}.

In the simplified kinematic circumstances where only 
a single representative momentum $q^2$ is considered, a basic model describing qualitatively
the resulting dynamics is given by 
\be
L(q^2) = b_0 + b_{\rm gl} \ln \left(\frac{q^2+m^2}{\Lambda^2}\right)
+ b_{\rm gh} \ln \left(\frac{q^2}{\Lambda^2}\right)\,,
\label{Lsupp}
\ee
where $L(q^2)$ denotes the particular combination of form factors, such that, at tree-level,
$L_0(q^2)=1$, and $b_0$, $b_{\rm gl}$, and $b_{\rm gh}$ are positive constants.

It is clear that, as \mbox{$q \to 0$}, the term with the unprotected logarithm will dominate over the others,
forcing $L(q^2)$ to reverse its sign (zero crossing), and finally diverge, \mbox{$L(0)\to -\infty$}.
Because, in practice,  $b_{\rm gl}$ is about one order of magnitude larger than $b_{\rm gh}$, 
the point where the unprotected logarithm overtakes the protected one is rather deep in the
infrared, and the location of the zero-crossing is at about $120$ MeV. Thus, in the intermediate region of momenta,
which is typically relevant for the onset of nonperturbative dynamics, we have \mbox{$L(q^2) < 1$};
this effect is known as the infrared suppression of the three-gluon vertex.

\section{\label{sec:pheno} Phenomenology}

In this section we discuss two representative phenomenological applications,  where the infrared suppression of the
corresponding form factors plays a crucial role. 

\subsection{\label{sec:pheno1} Effective couplings}

A typical quantity employed in a variety of phenomenological applications is
the effective charge, defined as a special {\it renormalization-group invariant}  combination
of propagators and vertex form factors.  In the case of the three-gluon vertex in the soft-gluon limit,
the corresponding charge, to be denoted by ${\alpha}_{\rm{3g}}(q^2)$, is
defined as~\cite{Alkofer:2004it,Fischer:2006ub,Athenodorou:2016oyh,Fu:2019hdw,Aguilar:2021okw} 
\be
{\alpha}_{\rm{3g}}(q^2)={\alpha}_s(\mu^2) \Lasym^2(q^2) \Dr^3(q^2) \,,
\label{coup_3g}
\ee
with $\Dr(q^2)$ defined in \1eq{eq:prop}.

It is natural to expect that
the infrared suppression of $\Lasym^2(q^2)$ will affect the shape and size of ${\alpha}_{\rm{3g}}(q^2)$.  
In order to meaningfully quantify this suppression, we compare ${\alpha}_{\rm{3g}}(q^2)$ with the
corresponding quantity defined from the ghost-gluon vertex, to be
denoted ${\alpha}_{\rm{cg}}(q^2)$, namely \mbox{(see, \eg~\cite{Alkofer:2004it, Fischer:2006ub,Aguilar:2009nf})}
\be
   {\alpha}_{\rm{cg}}(q^2) = {\alpha}_s(\mu^2) B_{1\,sg}^{2}(q^2) F^2(q^2)\Dr(q^2)\,,  
\label{coup_cg}
\ee
where $B_{1\,sg}(q^2)$ is the ghost-gluon form factor introduced in Fig.~\ref{fig:supvert}. 

\begin{figure}[h]
\centering
\includegraphics[scale=0.25]{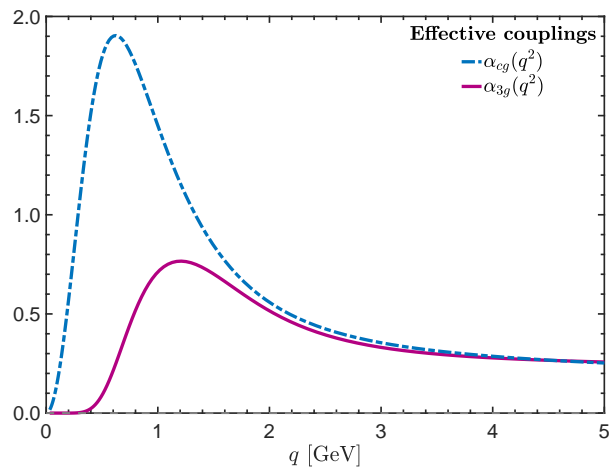}
\caption{ The comparison of the effective couplings, ${\alpha}_{\rm{cg}}(q^2)$ (blue dashed line) and ${\alpha}_{\rm{3g}}(q^2)$   (magenta continuous curve).}
\label{fig:coupling}
\end{figure}

It is important to mention that both effective couplings
are computed in the same renormalization scheme,  namely the Taylor scheme~\cite{Boucaud:2008gn,Boucaud:2011eh,vonSmekal:2009ae} where we have fixed  that \mbox{${\alpha}_s(\mu) = 0.244$},  at \mbox{$\mu = 4.3$  GeV} (for more details see~\cite{Aguilar:2021okw}).

The comparison of the two effective charges is displayed in Fig.~\ref{fig:coupling}.   One clearly  sees  that, as  
the momentum $q$ decreases,   ${\alpha}_{\rm{3g}}(q^2)$  (magenta continuous)   becomes considerably smaller than  ${\alpha}_{\rm{cg}}(q^2)$
  (blue dashed line).    The suppression of  ${\alpha}_{\rm{3g}}(q^2)$,    located in the region below \mbox{$2$ GeV} is consistent with previous finding~\cite{Huber:2012kd,Blum:2014gna,Williams:2014iea,Cyrol:2016tym,Cyrol:2017ewj, Aguilar:2020yni}, and its origin is exclusively associated  with the suppression of  the $\Lasym(q^2)$.

\subsection{\label{sec:pheno2} Pseudoscalar glueball}

The dynamical generation of a mass gap in pure-gauge QCD is intimately connected with the 
attendant appearance of glueball bound-states~\cite{Cornwall:1981zr}.  The rich glueball spectrum, and related fundamental properties,  
has been obtained by means of detailed lattice simulations, see \eg~\cite{Morningstar:1999rf,Bali:1993fb,McNeile:2008sr,Chen:2005mg,Athenodorou:2020ani}. Evidently, these results serve as valuable benchmarks in the
ongoing effort of continuum bound-state methods to reach an intuitive
understanding of the underlying dynamics~\cite{Dudal:2010cd,Meyers:2012ka, Souza:2019ylx,Huber:2021yfy}.

\begin{figure}[h]
\centering
\vspace{-0.25cm}
\includegraphics[scale=0.45]{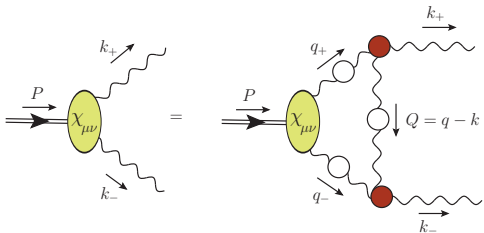}
\vspace{-0.25cm}
\caption{The diagrammatic representation of the Bethe-Salpeter equation for the pseudoscalar glueball with total momentum $P$,  and $\ell_{\pm} := P/2 \pm \ell$,  for  $\ell=k, q$ .}
\label{fig:BSE}
\end{figure}

In this context, the $J^{PC}=0^{-+}$ glueball represents
the simplest case, because the pertinent Bethe-Salpeter equation
possesses a single dynamical kernel, which essentially describes the four-gluon scattering process.
The lowest-order contribution of this kernel is shown in Fig.~\ref{fig:BSE}; evidently,
the three-gluon vertex constitutes one of its central ingredients~\cite{Souza:2019ylx}.

Moreover, 
the corresponding amplitude involves only one scalar function, namely
\begin{equation}
\chi_{\mu \nu}(k_+,k_-) = \epsilon_{\mu \nu \alpha \beta}k^{\alpha}P^{\beta}\mathcal{F}(k;P) \,,
\label{glupseudo}
\end{equation}
simplifying considerably the treatment of this problem.
%

It turns out that the infrared suppression of the three-gluon vertex, and the overall attenuation of the
interaction strength that it induces is
instrumental for the formation of the pseudoscalar glueball state, with a mass compatible with that
obtained from the lattice~\cite{Souza:2019ylx}.

Let us finally mention that 
the need for  a considerable suppression has also been established   
in studies of hybrid states by means of Faddeev equations~\cite{Xu:2018cor}.

\section{\label{sec:conc}Conclusions}
\vspace{-0.25cm}

In this presentation we have reviewed some of the most characteristic nonpertubative features
of the three-gluon vertex, unraveled by the ongoing synergy of a 
multitude of techniques and approaches, such as functional methods, lattice simulations, and
STI-based constructions.

We have focused on the interplay between the dynamics of the three-gluon vertex and the
Landau-gauge two-point sector of the theory.
In particular, as has been argued in Sec.~\ref{sec:supp}, the characteristic infrared suppression displayed
by the main form factors of the three-gluon vertex is tightly interlocked with the
mass generating pattern established in the gauge sector of QCD. 

There is an additional key aspect of the three-gluon vertex, which is worth mentioning, albeit in passing.
In particular, the three-gluon vertex develops {\it longitudinally coupled bound state massless poles}, which trigger
the well-known Schwinger mechanism~\cite{Schwinger:1962tn,Schwinger:1962tp}, endowing the gluons with a dynamical mass scale~\cite{Aguilar:2008xm,Binosi:2009qm,Binosi:2009qm}.  Due to their
special kinematic properties, these poles decouple from the transversally projected vertex $\overline{\Gamma}_{\alpha \mu \nu}(q,r,p)$ [see \1eq{eq:Gammabar}], which enters 
in the lattice quantities defined according to \1eq{eq:latobs}. Consequently, these dynamically produced poles  
do not induce divergences in the results displayed in Fig.~\ref{fig:sym_asym} and Fig.~\ref{fig:unquenched}.
Nonetheless, as has been recently demonstrated in~\cite{Aguilar:2021uwa}, the massless poles leave
smoking-gun signals  of their presence, by inducing finite displacements to the non-Abelian Ward identity
satisfied by the pole-free part of the three-gluon vertex. Quite interestingly, this displacement
is identical to the Bethe-Salpeter amplitude that controls the dynamical formation of the
massless poles~\cite{Aguilar:2011xe,Ibanez:2012zk,Aguilar:2017dco,Binosi:2017rwj}, thus establishing a powerful constraint on the entire mass generating
mechanism put forth in a series of works (see~\cite{Aguilar:2021uwa} and references therein).

It would be clearly important to continue the research activity surrounding the three-gluon vertex
in the future. In this context, a major challenge for functional methods is the
extension of the results for this vertex from space-like to time-like momenta.
Such information will be particularly important, both from the theoretical as well as
the phenomenological point of view. 
The methods and techniques developed in~\cite{Cyrol:2018xeq,Horak:2020eng,Horak:2021pfr,Horak:2021syv} may be decisive
for making progress with this demanding endeavor.

\section*{Acknowledgments}
\label{sec:acknowledgments}
\vspace{-0.25cm}

We thank the organizers of the 19th International Conference on Hadron Spectroscopy and Structure (Hadron 21-virtual) for the kind invitation.
The work of J.~P. is supported by the  Spanish AEI-MICINN grant PID2020-113334GB-I00/AEI/10.13039/501100011033,
and the  grant  Prometeo/2019/087 of the Generalitat Valenciana.  A.~C.~A. is supported by the  CNPq grants 307854/2019-1 and 464898/2014-5 (INCT-FNA).
A.~C.~A. and M.~N.~F.  also acknowledge financial support from  the FAPESP projects 2017/05685-2 and 2020/12795-1, respectively.



\begin{thebibliography}{85}%
\makeatletter
\providecommand \@ifxundefined [1]{%
 \@ifx{#1\undefined}
}%
\providecommand \@ifnum [1]{%
 \ifnum #1\expandafter \@firstoftwo
 \else \expandafter \@secondoftwo
 \fi
}%
\providecommand \@ifx [1]{%
 \ifx #1\expandafter \@firstoftwo
 \else \expandafter \@secondoftwo
 \fi
}%
\providecommand \natexlab [1]{#1}%
\providecommand \enquote  [1]{``#1''}%
\providecommand \bibnamefont  [1]{#1}%
\providecommand \bibfnamefont [1]{#1}%
\providecommand \citenamefont [1]{#1}%
\providecommand \href@noop [0]{\@secondoftwo}%
\providecommand \href [0]{\begingroup \@sanitize@url \@href}%
\providecommand \@href[1]{\@@startlink{#1}\@@href}%
\providecommand \@@href[1]{\endgroup#1\@@endlink}%
\providecommand \@sanitize@url [0]{\catcode `\\12\catcode `\$12\catcode
  `\&12\catcode `\#12\catcode `\^12\catcode `\_12\catcode `\%12\relax}%
\providecommand \@@startlink[1]{}%
\providecommand \@@endlink[0]{}%
\providecommand \url  [0]{\begingroup\@sanitize@url \@url }%
\providecommand \@url [1]{\endgroup\@href {#1}{\urlprefix }}%
\providecommand \urlprefix  [0]{URL }%
\providecommand \Eprint [0]{\href }%
\providecommand \doibase [0]{http://dx.doi.org/}%
\providecommand \selectlanguage [0]{\@gobble}%
\providecommand \bibinfo  [0]{\@secondoftwo}%
\providecommand \bibfield  [0]{\@secondoftwo}%
\providecommand \translation [1]{[#1]}%
\providecommand \BibitemOpen [0]{}%
\providecommand \bibitemStop [0]{}%
\providecommand \bibitemNoStop [0]{.\EOS\space}%
\providecommand \EOS [0]{\spacefactor3000\relax}%
\providecommand \BibitemShut  [1]{\csname bibitem#1\endcsname}%
\let\auto@bib@innerbib\@empty
\bibitem [{\citenamefont {Marciano}\ and\ \citenamefont
  {Pagels}(1978)}]{Marciano:1977su}%
  \BibitemOpen
  \bibfield  {author} {\bibinfo {author} {\bibfnamefont {W.~J.}\ \bibnamefont
  {Marciano}}\ and\ \bibinfo {author} {\bibfnamefont {H.}~\bibnamefont
  {Pagels}},\ }\href {\doibase 10.1016/0370-1573(78)90208-9} {\bibfield
  {journal} {\bibinfo  {journal} {Phys. Rept.}\ }\textbf {\bibinfo {volume}
  {36}},\ \bibinfo {pages} {137} (\bibinfo {year} {1978})}\BibitemShut
  {NoStop}%
\bibitem [{\citenamefont {Ball}\ and\ \citenamefont
  {Chiu}(1980)}]{Ball:1980ax}%
  \BibitemOpen
  \bibfield  {author} {\bibinfo {author} {\bibfnamefont {J.~S.}\ \bibnamefont
  {Ball}}\ and\ \bibinfo {author} {\bibfnamefont {T.-W.}\ \bibnamefont
  {Chiu}},\ }\href {\doibase 10.1103/PhysRevD.22.2550} {\bibfield  {journal}
  {\bibinfo  {journal} {Phys. Rev. D}\ }\textbf {\bibinfo {volume} {22}},\
  \bibinfo {pages} {2550} (\bibinfo {year} {1980})},\ \bibinfo {note}
  {[Erratum: Phys.Rev.D 23, 3085 (1981)]}\BibitemShut {NoStop}%
\bibitem [{\citenamefont {Davydychev}\ \emph {et~al.}(1996)\citenamefont
  {Davydychev}, \citenamefont {Osland},\ and\ \citenamefont
  {Tarasov}}]{Davydychev:1996pb}%
  \BibitemOpen
  \bibfield  {author} {\bibinfo {author} {\bibfnamefont {A.~I.}\ \bibnamefont
  {Davydychev}}, \bibinfo {author} {\bibfnamefont {P.}~\bibnamefont {Osland}},
  \ and\ \bibinfo {author} {\bibfnamefont {O.}~\bibnamefont {Tarasov}},\ }\href
  {\doibase 10.1103/PhysRevD.59.109901} {\bibfield  {journal} {\bibinfo
  {journal} {Phys. Rev. D}\ }\textbf {\bibinfo {volume} {54}},\ \bibinfo
  {pages} {4087} (\bibinfo {year} {1996})},\ \bibinfo {note} {[Erratum:
  Phys.Rev.D 59, 109901 (1999)]}\BibitemShut {NoStop}%
\bibitem [{\citenamefont {Gross}\ and\ \citenamefont
  {Wilczek}(1973)}]{Gross:1973id}%
  \BibitemOpen
  \bibfield  {author} {\bibinfo {author} {\bibfnamefont {D.~J.}\ \bibnamefont
  {Gross}}\ and\ \bibinfo {author} {\bibfnamefont {F.}~\bibnamefont
  {Wilczek}},\ }\href {\doibase 10.1103/PhysRevLett.30.1343} {\bibfield
  {journal} {\bibinfo  {journal} {Phys. Rev. Lett.}\ }\textbf {\bibinfo
  {volume} {30}},\ \bibinfo {pages} {1343} (\bibinfo {year}
  {1973})}\BibitemShut {NoStop}%
\bibitem [{\citenamefont {Politzer}(1973)}]{Politzer:1973fx}%
  \BibitemOpen
  \bibfield  {author} {\bibinfo {author} {\bibfnamefont {H.~D.}\ \bibnamefont
  {Politzer}},\ }\href {\doibase 10.1103/PhysRevLett.30.1346} {\bibfield
  {journal} {\bibinfo  {journal} {Phys. Rev. Lett.}\ }\textbf {\bibinfo
  {volume} {30}},\ \bibinfo {pages} {1346} (\bibinfo {year}
  {1973})}\BibitemShut {NoStop}%
\bibitem [{\citenamefont {Alkofer}\ \emph {et~al.}(2005)\citenamefont
  {Alkofer}, \citenamefont {Fischer},\ and\ \citenamefont
  {Llanes-Estrada}}]{Alkofer:2004it}%
  \BibitemOpen
  \bibfield  {author} {\bibinfo {author} {\bibfnamefont {R.}~\bibnamefont
  {Alkofer}}, \bibinfo {author} {\bibfnamefont {C.~S.}\ \bibnamefont
  {Fischer}}, \ and\ \bibinfo {author} {\bibfnamefont {F.~J.}\ \bibnamefont
  {Llanes-Estrada}},\ }\href {\doibase 10.1016/j.physletb.2008.11.068}
  {\bibfield  {journal} {\bibinfo  {journal} {Phys. Lett. B}\ }\textbf
  {\bibinfo {volume} {611}},\ \bibinfo {pages} {279} (\bibinfo {year}
  {2005})},\ \bibinfo {note} {[Erratum: Phys.Lett.B 670, 460--461
  (2009)]}\BibitemShut {NoStop}%
\bibitem [{\citenamefont {Huber}\ \emph {et~al.}(2012)\citenamefont {Huber},
  \citenamefont {Maas},\ and\ \citenamefont {von Smekal}}]{Huber:2012zj}%
  \BibitemOpen
  \bibfield  {author} {\bibinfo {author} {\bibfnamefont {M.~Q.}\ \bibnamefont
  {Huber}}, \bibinfo {author} {\bibfnamefont {A.}~\bibnamefont {Maas}}, \ and\
  \bibinfo {author} {\bibfnamefont {L.}~\bibnamefont {von Smekal}},\ }\href
  {\doibase 10.1007/JHEP11(2012)035} {\bibfield  {journal} {\bibinfo  {journal}
  {J. High Energy Phys.}\ }\textbf {\bibinfo {volume} {11}},\ \bibinfo {pages}
  {035} (\bibinfo {year} {2012})}\BibitemShut {NoStop}%
\bibitem [{\citenamefont {Pelaez}\ \emph {et~al.}(2013)\citenamefont {Pelaez},
  \citenamefont {Tissier},\ and\ \citenamefont {Wschebor}}]{Pelaez:2013cpa}%
  \BibitemOpen
  \bibfield  {author} {\bibinfo {author} {\bibfnamefont {M.}~\bibnamefont
  {Pelaez}}, \bibinfo {author} {\bibfnamefont {M.}~\bibnamefont {Tissier}}, \
  and\ \bibinfo {author} {\bibfnamefont {N.}~\bibnamefont {Wschebor}},\ }\href
  {\doibase 10.1103/PhysRevD.88.125003} {\bibfield  {journal} {\bibinfo
  {journal} {Phys. Rev.}\ }\textbf {\bibinfo {volume} {D88}},\ \bibinfo {pages}
  {125003} (\bibinfo {year} {2013})}\BibitemShut {NoStop}%
\bibitem [{\citenamefont {Aguilar}\ \emph {et~al.}(2014)\citenamefont
  {Aguilar}, \citenamefont {Binosi}, \citenamefont {Iba{\~n}ez},\ and\
  \citenamefont {Papavassiliou}}]{Aguilar:2013vaa}%
  \BibitemOpen
  \bibfield  {author} {\bibinfo {author} {\bibfnamefont {A.~C.}\ \bibnamefont
  {Aguilar}}, \bibinfo {author} {\bibfnamefont {D.}~\bibnamefont {Binosi}},
  \bibinfo {author} {\bibfnamefont {D.}~\bibnamefont {Iba{\~n}ez}}, \ and\
  \bibinfo {author} {\bibfnamefont {J.}~\bibnamefont {Papavassiliou}},\ }\href
  {\doibase 10.1103/PhysRevD.89.085008} {\bibfield  {journal} {\bibinfo
  {journal} {Phys. Rev.}\ }\textbf {\bibinfo {volume} {D89}},\ \bibinfo {pages}
  {085008} (\bibinfo {year} {2014})}\BibitemShut {NoStop}%
\bibitem [{\citenamefont {Blum}\ \emph {et~al.}(2014)\citenamefont {Blum},
  \citenamefont {Huber}, \citenamefont {Mitter},\ and\ \citenamefont {von
  Smekal}}]{Blum:2014gna}%
  \BibitemOpen
  \bibfield  {author} {\bibinfo {author} {\bibfnamefont {A.}~\bibnamefont
  {Blum}}, \bibinfo {author} {\bibfnamefont {M.~Q.}\ \bibnamefont {Huber}},
  \bibinfo {author} {\bibfnamefont {M.}~\bibnamefont {Mitter}}, \ and\ \bibinfo
  {author} {\bibfnamefont {L.}~\bibnamefont {von Smekal}},\ }\href {\doibase
  10.1103/PhysRevD.89.061703} {\bibfield  {journal} {\bibinfo  {journal} {Phys.
  Rev.}\ }\textbf {\bibinfo {volume} {D89}},\ \bibinfo {pages} {061703}
  (\bibinfo {year} {2014})}\BibitemShut {NoStop}%
\bibitem [{\citenamefont {Eichmann}\ \emph {et~al.}(2014)\citenamefont
  {Eichmann}, \citenamefont {Williams}, \citenamefont {Alkofer},\ and\
  \citenamefont {Vujinovic}}]{Eichmann:2014xya}%
  \BibitemOpen
  \bibfield  {author} {\bibinfo {author} {\bibfnamefont {G.}~\bibnamefont
  {Eichmann}}, \bibinfo {author} {\bibfnamefont {R.}~\bibnamefont {Williams}},
  \bibinfo {author} {\bibfnamefont {R.}~\bibnamefont {Alkofer}}, \ and\
  \bibinfo {author} {\bibfnamefont {M.}~\bibnamefont {Vujinovic}},\ }\href
  {\doibase 10.1103/PhysRevD.89.105014} {\bibfield  {journal} {\bibinfo
  {journal} {Phys. Rev.}\ }\textbf {\bibinfo {volume} {D89}},\ \bibinfo {pages}
  {105014} (\bibinfo {year} {2014})}\BibitemShut {NoStop}%
\bibitem [{\citenamefont {Mitter}\ \emph {et~al.}(2015)\citenamefont {Mitter},
  \citenamefont {Pawlowski},\ and\ \citenamefont
  {Strodthoff}}]{Mitter:2014wpa}%
  \BibitemOpen
  \bibfield  {author} {\bibinfo {author} {\bibfnamefont {M.}~\bibnamefont
  {Mitter}}, \bibinfo {author} {\bibfnamefont {J.~M.}\ \bibnamefont
  {Pawlowski}}, \ and\ \bibinfo {author} {\bibfnamefont {N.}~\bibnamefont
  {Strodthoff}},\ }\href {\doibase 10.1103/PhysRevD.91.054035} {\bibfield
  {journal} {\bibinfo  {journal} {Phys. Rev.}\ }\textbf {\bibinfo {volume}
  {D91}},\ \bibinfo {pages} {054035} (\bibinfo {year} {2015})}\BibitemShut
  {NoStop}%
\bibitem [{\citenamefont {Williams}\ \emph {et~al.}(2016)\citenamefont
  {Williams}, \citenamefont {Fischer},\ and\ \citenamefont
  {Heupel}}]{Williams:2015cvx}%
  \BibitemOpen
  \bibfield  {author} {\bibinfo {author} {\bibfnamefont {R.}~\bibnamefont
  {Williams}}, \bibinfo {author} {\bibfnamefont {C.~S.}\ \bibnamefont
  {Fischer}}, \ and\ \bibinfo {author} {\bibfnamefont {W.}~\bibnamefont
  {Heupel}},\ }\href {\doibase 10.1103/PhysRevD.93.034026} {\bibfield
  {journal} {\bibinfo  {journal} {Phys. Rev.}\ }\textbf {\bibinfo {volume}
  {D93}},\ \bibinfo {pages} {034026} (\bibinfo {year} {2016})}\BibitemShut
  {NoStop}%
\bibitem [{\citenamefont {Blum}\ \emph {et~al.}(2015)\citenamefont {Blum},
  \citenamefont {Alkofer}, \citenamefont {Huber},\ and\ \citenamefont
  {Windisch}}]{Blum:2015lsa}%
  \BibitemOpen
  \bibfield  {author} {\bibinfo {author} {\bibfnamefont {A.~L.}\ \bibnamefont
  {Blum}}, \bibinfo {author} {\bibfnamefont {R.}~\bibnamefont {Alkofer}},
  \bibinfo {author} {\bibfnamefont {M.~Q.}\ \bibnamefont {Huber}}, \ and\
  \bibinfo {author} {\bibfnamefont {A.}~\bibnamefont {Windisch}},\ }\href
  {\doibase 10.5506/APhysPolBSupp.8.321} {\bibfield  {journal} {\bibinfo
  {journal} {Acta Phys. Polon. Supp.}\ }\textbf {\bibinfo {volume} {8}},\
  \bibinfo {pages} {321} (\bibinfo {year} {2015})}\BibitemShut {NoStop}%
\bibitem [{\citenamefont {Cyrol}\ \emph {et~al.}(2016)\citenamefont {Cyrol},
  \citenamefont {Fister}, \citenamefont {Mitter}, \citenamefont {Pawlowski},\
  and\ \citenamefont {Strodthoff}}]{Cyrol:2016tym}%
  \BibitemOpen
  \bibfield  {author} {\bibinfo {author} {\bibfnamefont {A.~K.}\ \bibnamefont
  {Cyrol}}, \bibinfo {author} {\bibfnamefont {L.}~\bibnamefont {Fister}},
  \bibinfo {author} {\bibfnamefont {M.}~\bibnamefont {Mitter}}, \bibinfo
  {author} {\bibfnamefont {J.~M.}\ \bibnamefont {Pawlowski}}, \ and\ \bibinfo
  {author} {\bibfnamefont {N.}~\bibnamefont {Strodthoff}},\ }\href {\doibase
  10.1103/PhysRevD.94.054005} {\bibfield  {journal} {\bibinfo  {journal} {Phys.
  Rev.}\ }\textbf {\bibinfo {volume} {D94}},\ \bibinfo {pages} {054005}
  (\bibinfo {year} {2016})}\BibitemShut {NoStop}%
\bibitem [{\citenamefont {Corell}\ \emph {et~al.}(2018)\citenamefont {Corell},
  \citenamefont {Cyrol}, \citenamefont {Mitter}, \citenamefont {Pawlowski},\
  and\ \citenamefont {Strodthoff}}]{Corell:2018yil}%
  \BibitemOpen
  \bibfield  {author} {\bibinfo {author} {\bibfnamefont {L.}~\bibnamefont
  {Corell}}, \bibinfo {author} {\bibfnamefont {A.~K.}\ \bibnamefont {Cyrol}},
  \bibinfo {author} {\bibfnamefont {M.}~\bibnamefont {Mitter}}, \bibinfo
  {author} {\bibfnamefont {J.~M.}\ \bibnamefont {Pawlowski}}, \ and\ \bibinfo
  {author} {\bibfnamefont {N.}~\bibnamefont {Strodthoff}},\ }\href {\doibase
  10.21468/SciPostPhys.5.6.066} {\bibfield  {journal} {\bibinfo  {journal}
  {SciPost Phys.}\ }\textbf {\bibinfo {volume} {5}},\ \bibinfo {pages} {066}
  (\bibinfo {year} {2018})}\BibitemShut {NoStop}%
\bibitem [{\citenamefont {Boucaud}\ \emph {et~al.}(2017)\citenamefont
  {Boucaud}, \citenamefont {De~Soto}, \citenamefont {Rodr\'{\i}guez-Quintero},\
  and\ \citenamefont {Zafeiropoulos}}]{Boucaud:2017obn}%
  \BibitemOpen
  \bibfield  {author} {\bibinfo {author} {\bibfnamefont {P.}~\bibnamefont
  {Boucaud}}, \bibinfo {author} {\bibfnamefont {F.}~\bibnamefont {De~Soto}},
  \bibinfo {author} {\bibfnamefont {J.}~\bibnamefont
  {Rodr\'{\i}guez-Quintero}}, \ and\ \bibinfo {author} {\bibfnamefont
  {S.}~\bibnamefont {Zafeiropoulos}},\ }\href {\doibase
  10.1103/PhysRevD.95.114503} {\bibfield  {journal} {\bibinfo  {journal} {Phys.
  Rev.}\ }\textbf {\bibinfo {volume} {D95}},\ \bibinfo {pages} {114503}
  (\bibinfo {year} {2017})}\BibitemShut {NoStop}%
\bibitem [{\citenamefont {Huber}(2020)}]{Huber:2018ned}%
  \BibitemOpen
  \bibfield  {author} {\bibinfo {author} {\bibfnamefont {M.~Q.}\ \bibnamefont
  {Huber}},\ }\href {\doibase 10.1016/j.physrep.2020.04.004} {\bibfield
  {journal} {\bibinfo  {journal} {Phys. Rept.}\ }\textbf {\bibinfo {volume}
  {879}},\ \bibinfo {pages} {1} (\bibinfo {year} {2020})}\BibitemShut {NoStop}%
\bibitem [{\citenamefont {Aguilar}\ \emph
  {et~al.}(2019{\natexlab{a}})\citenamefont {Aguilar}, \citenamefont
  {Ferreira}, \citenamefont {Figueiredo},\ and\ \citenamefont
  {Papavassiliou}}]{Aguilar:2019jsj}%
  \BibitemOpen
  \bibfield  {author} {\bibinfo {author} {\bibfnamefont {A.~C.}\ \bibnamefont
  {Aguilar}}, \bibinfo {author} {\bibfnamefont {M.~N.}\ \bibnamefont
  {Ferreira}}, \bibinfo {author} {\bibfnamefont {C.~T.}\ \bibnamefont
  {Figueiredo}}, \ and\ \bibinfo {author} {\bibfnamefont {J.}~\bibnamefont
  {Papavassiliou}},\ }\href {\doibase 10.1103/PhysRevD.99.094010} {\bibfield
  {journal} {\bibinfo  {journal} {Phys. Rev.}\ }\textbf {\bibinfo {volume}
  {D99}},\ \bibinfo {pages} {094010} (\bibinfo {year}
  {2019}{\natexlab{a}})}\BibitemShut {NoStop}%
\bibitem [{\citenamefont {Aguilar}\ \emph
  {et~al.}(2020{\natexlab{a}})\citenamefont {Aguilar}, \citenamefont {De~Soto},
  \citenamefont {Ferreira}, \citenamefont {Papavassiliou}, \citenamefont
  {Rodr\'\i{}guez-Quintero},\ and\ \citenamefont
  {Zafeiropoulos}}]{Aguilar:2019uob}%
  \BibitemOpen
  \bibfield  {author} {\bibinfo {author} {\bibfnamefont {A.~C.}\ \bibnamefont
  {Aguilar}}, \bibinfo {author} {\bibfnamefont {F.}~\bibnamefont {De~Soto}},
  \bibinfo {author} {\bibfnamefont {M.~N.}\ \bibnamefont {Ferreira}}, \bibinfo
  {author} {\bibfnamefont {J.}~\bibnamefont {Papavassiliou}}, \bibinfo {author}
  {\bibfnamefont {J.}~\bibnamefont {Rodr\'\i{}guez-Quintero}}, \ and\ \bibinfo
  {author} {\bibfnamefont {S.}~\bibnamefont {Zafeiropoulos}},\ }\href {\doibase
  10.1140/epjc/s10052-020-7741-0} {\bibfield  {journal} {\bibinfo  {journal}
  {Eur. Phys. J. C}\ }\textbf {\bibinfo {volume} {80}},\ \bibinfo {pages} {154}
  (\bibinfo {year} {2020}{\natexlab{a}})}\BibitemShut {NoStop}%
\bibitem [{\citenamefont {Aguilar}\ \emph
  {et~al.}(2019{\natexlab{b}})\citenamefont {Aguilar}, \citenamefont
  {Ferreira}, \citenamefont {Figueiredo},\ and\ \citenamefont
  {Papavassiliou}}]{Aguilar:2019kxz}%
  \BibitemOpen
  \bibfield  {author} {\bibinfo {author} {\bibfnamefont {A.~C.}\ \bibnamefont
  {Aguilar}}, \bibinfo {author} {\bibfnamefont {M.~N.}\ \bibnamefont
  {Ferreira}}, \bibinfo {author} {\bibfnamefont {C.~T.}\ \bibnamefont
  {Figueiredo}}, \ and\ \bibinfo {author} {\bibfnamefont {J.}~\bibnamefont
  {Papavassiliou}},\ }\href {\doibase 10.1103/PhysRevD.100.094039} {\bibfield
  {journal} {\bibinfo  {journal} {Phys. Rev. D}\ }\textbf {\bibinfo {volume}
  {100}},\ \bibinfo {pages} {094039} (\bibinfo {year}
  {2019}{\natexlab{b}})}\BibitemShut {NoStop}%
\bibitem [{\citenamefont {Parrinello}(1994)}]{Parrinello:1994wd}%
  \BibitemOpen
  \bibfield  {author} {\bibinfo {author} {\bibfnamefont {C.}~\bibnamefont
  {Parrinello}},\ }\href {\doibase 10.1103/PhysRevD.50.R4247} {\bibfield
  {journal} {\bibinfo  {journal} {Phys. Rev.}\ }\textbf {\bibinfo {volume}
  {D50}},\ \bibinfo {pages} {R4247} (\bibinfo {year} {1994})}\BibitemShut
  {NoStop}%
\bibitem [{\citenamefont {Alles}\ \emph {et~al.}(1997)\citenamefont {Alles},
  \citenamefont {Henty}, \citenamefont {Panagopoulos}, \citenamefont
  {Parrinello}, \citenamefont {Pittori},\ and\ \citenamefont
  {Richards}}]{Alles:1996ka}%
  \BibitemOpen
  \bibfield  {author} {\bibinfo {author} {\bibfnamefont {B.}~\bibnamefont
  {Alles}}, \bibinfo {author} {\bibfnamefont {D.}~\bibnamefont {Henty}},
  \bibinfo {author} {\bibfnamefont {H.}~\bibnamefont {Panagopoulos}}, \bibinfo
  {author} {\bibfnamefont {C.}~\bibnamefont {Parrinello}}, \bibinfo {author}
  {\bibfnamefont {C.}~\bibnamefont {Pittori}}, \ and\ \bibinfo {author}
  {\bibfnamefont {D.~G.}\ \bibnamefont {Richards}},\ }\href {\doibase
  10.1016/S0550-3213(97)00483-5} {\bibfield  {journal} {\bibinfo  {journal}
  {Nucl. Phys.}\ }\textbf {\bibinfo {volume} {B502}},\ \bibinfo {pages} {325}
  (\bibinfo {year} {1997})}\BibitemShut {NoStop}%
\bibitem [{\citenamefont {Parrinello}\ \emph {et~al.}(1998)\citenamefont
  {Parrinello}, \citenamefont {Richards}, \citenamefont {Alles}, \citenamefont
  {Panagopoulos},\ and\ \citenamefont {Pittori}}]{Parrinello:1997wm}%
  \BibitemOpen
  \bibfield  {author} {\bibinfo {author} {\bibfnamefont {C.}~\bibnamefont
  {Parrinello}}, \bibinfo {author} {\bibfnamefont {D.}~\bibnamefont
  {Richards}}, \bibinfo {author} {\bibfnamefont {B.}~\bibnamefont {Alles}},
  \bibinfo {author} {\bibfnamefont {H.}~\bibnamefont {Panagopoulos}}, \ and\
  \bibinfo {author} {\bibfnamefont {C.}~\bibnamefont {Pittori}} (\bibinfo
  {collaboration} {UKQCD}),\ }\href {\doibase 10.1016/S0920-5632(97)00734-2}
  {\bibfield  {journal} {\bibinfo  {journal} {Nucl. Phys. B Proc. Suppl.}\
  }\textbf {\bibinfo {volume} {63}},\ \bibinfo {pages} {245} (\bibinfo {year}
  {1998})}\BibitemShut {NoStop}%
\bibitem [{\citenamefont {Boucaud}\ \emph {et~al.}(1998)\citenamefont
  {Boucaud}, \citenamefont {Leroy}, \citenamefont {Micheli}, \citenamefont
  {Pene},\ and\ \citenamefont {Roiesnel}}]{Boucaud:1998bq}%
  \BibitemOpen
  \bibfield  {author} {\bibinfo {author} {\bibfnamefont {P.}~\bibnamefont
  {Boucaud}}, \bibinfo {author} {\bibfnamefont {J.~P.}\ \bibnamefont {Leroy}},
  \bibinfo {author} {\bibfnamefont {J.}~\bibnamefont {Micheli}}, \bibinfo
  {author} {\bibfnamefont {O.}~\bibnamefont {Pene}}, \ and\ \bibinfo {author}
  {\bibfnamefont {C.}~\bibnamefont {Roiesnel}},\ }\href {\doibase
  10.1088/1126-6708/1998/10/017} {\bibfield  {journal} {\bibinfo  {journal} {J.
  High Energy Phys.}\ }\textbf {\bibinfo {volume} {10}},\ \bibinfo {pages}
  {017} (\bibinfo {year} {1998})}\BibitemShut {NoStop}%
\bibitem [{\citenamefont {Cucchieri}\ \emph {et~al.}(2006)\citenamefont
  {Cucchieri}, \citenamefont {Maas},\ and\ \citenamefont
  {Mendes}}]{Cucchieri:2006tf}%
  \BibitemOpen
  \bibfield  {author} {\bibinfo {author} {\bibfnamefont {A.}~\bibnamefont
  {Cucchieri}}, \bibinfo {author} {\bibfnamefont {A.}~\bibnamefont {Maas}}, \
  and\ \bibinfo {author} {\bibfnamefont {T.}~\bibnamefont {Mendes}},\ }\href
  {\doibase 10.1103/PhysRevD.74.014503} {\bibfield  {journal} {\bibinfo
  {journal} {Phys. Rev.}\ }\textbf {\bibinfo {volume} {D74}},\ \bibinfo {pages}
  {014503} (\bibinfo {year} {2006})}\BibitemShut {NoStop}%
\bibitem [{\citenamefont {Cucchieri}\ \emph {et~al.}(2008)\citenamefont
  {Cucchieri}, \citenamefont {Maas},\ and\ \citenamefont
  {Mendes}}]{Cucchieri:2008qm}%
  \BibitemOpen
  \bibfield  {author} {\bibinfo {author} {\bibfnamefont {A.}~\bibnamefont
  {Cucchieri}}, \bibinfo {author} {\bibfnamefont {A.}~\bibnamefont {Maas}}, \
  and\ \bibinfo {author} {\bibfnamefont {T.}~\bibnamefont {Mendes}},\ }\href
  {\doibase 10.1103/PhysRevD.77.094510} {\bibfield  {journal} {\bibinfo
  {journal} {Phys. Rev.}\ }\textbf {\bibinfo {volume} {D77}},\ \bibinfo {pages}
  {094510} (\bibinfo {year} {2008})}\BibitemShut {NoStop}%
\bibitem [{\citenamefont {Athenodorou}\ \emph {et~al.}(2016)\citenamefont
  {Athenodorou}, \citenamefont {Binosi}, \citenamefont {Boucaud}, \citenamefont
  {De~Soto}, \citenamefont {Papavassiliou}, \citenamefont
  {Rodriguez-Quintero},\ and\ \citenamefont
  {Zafeiropoulos}}]{Athenodorou:2016oyh}%
  \BibitemOpen
  \bibfield  {author} {\bibinfo {author} {\bibfnamefont {A.}~\bibnamefont
  {Athenodorou}}, \bibinfo {author} {\bibfnamefont {D.}~\bibnamefont {Binosi}},
  \bibinfo {author} {\bibfnamefont {P.}~\bibnamefont {Boucaud}}, \bibinfo
  {author} {\bibfnamefont {F.}~\bibnamefont {De~Soto}}, \bibinfo {author}
  {\bibfnamefont {J.}~\bibnamefont {Papavassiliou}}, \bibinfo {author}
  {\bibfnamefont {J.}~\bibnamefont {Rodriguez-Quintero}}, \ and\ \bibinfo
  {author} {\bibfnamefont {S.}~\bibnamefont {Zafeiropoulos}},\ }\href {\doibase
  10.1016/j.physletb.2016.08.065} {\bibfield  {journal} {\bibinfo  {journal}
  {Phys. Lett.}\ }\textbf {\bibinfo {volume} {B761}},\ \bibinfo {pages} {444}
  (\bibinfo {year} {2016})}\BibitemShut {NoStop}%
\bibitem [{\citenamefont {Duarte}\ \emph {et~al.}(2016)\citenamefont {Duarte},
  \citenamefont {Oliveira},\ and\ \citenamefont {Silva}}]{Duarte:2016ieu}%
  \BibitemOpen
  \bibfield  {author} {\bibinfo {author} {\bibfnamefont {A.~G.}\ \bibnamefont
  {Duarte}}, \bibinfo {author} {\bibfnamefont {O.}~\bibnamefont {Oliveira}}, \
  and\ \bibinfo {author} {\bibfnamefont {P.~J.}\ \bibnamefont {Silva}},\ }\href
  {\doibase 10.1103/PhysRevD.94.074502} {\bibfield  {journal} {\bibinfo
  {journal} {Phys. Rev.}\ }\textbf {\bibinfo {volume} {D94}},\ \bibinfo {pages}
  {074502} (\bibinfo {year} {2016})}\BibitemShut {NoStop}%
\bibitem [{\citenamefont {Vujinovic}\ and\ \citenamefont
  {Mendes}(2019)}]{Vujinovic:2018nqc}%
  \BibitemOpen
  \bibfield  {author} {\bibinfo {author} {\bibfnamefont {M.}~\bibnamefont
  {Vujinovic}}\ and\ \bibinfo {author} {\bibfnamefont {T.}~\bibnamefont
  {Mendes}},\ }\href {\doibase 10.1103/PhysRevD.99.034501} {\bibfield
  {journal} {\bibinfo  {journal} {Phys. Rev.}\ }\textbf {\bibinfo {volume}
  {D99}},\ \bibinfo {pages} {034501} (\bibinfo {year} {2019})}\BibitemShut
  {NoStop}%
\bibitem [{\citenamefont {Cornwall}(1982)}]{Cornwall:1981zr}%
  \BibitemOpen
  \bibfield  {author} {\bibinfo {author} {\bibfnamefont {J.~M.}\ \bibnamefont
  {Cornwall}},\ }\href {\doibase 10.1103/PhysRevD.26.1453} {\bibfield
  {journal} {\bibinfo  {journal} {Phys. Rev. D}\ }\textbf {\bibinfo {volume}
  {26}},\ \bibinfo {pages} {1453} (\bibinfo {year} {1982})}\BibitemShut
  {NoStop}%
\bibitem [{\citenamefont {Alkofer}\ \emph {et~al.}(2009)\citenamefont
  {Alkofer}, \citenamefont {Huber},\ and\ \citenamefont
  {Schwenzer}}]{Alkofer:2008dt}%
  \BibitemOpen
  \bibfield  {author} {\bibinfo {author} {\bibfnamefont {R.}~\bibnamefont
  {Alkofer}}, \bibinfo {author} {\bibfnamefont {M.~Q.}\ \bibnamefont {Huber}},
  \ and\ \bibinfo {author} {\bibfnamefont {K.}~\bibnamefont {Schwenzer}},\
  }\href {\doibase 10.1140/epjc/s10052-009-1066-3} {\bibfield  {journal}
  {\bibinfo  {journal} {Eur. Phys. J. C}\ }\textbf {\bibinfo {volume} {62}},\
  \bibinfo {pages} {761} (\bibinfo {year} {2009})}\BibitemShut {NoStop}%
\bibitem [{\citenamefont {Alkofer}\ \emph {et~al.}(2010)\citenamefont
  {Alkofer}, \citenamefont {Huber},\ and\ \citenamefont
  {Schwenzer}}]{Alkofer:2008jy}%
  \BibitemOpen
  \bibfield  {author} {\bibinfo {author} {\bibfnamefont {R.}~\bibnamefont
  {Alkofer}}, \bibinfo {author} {\bibfnamefont {M.~Q.}\ \bibnamefont {Huber}},
  \ and\ \bibinfo {author} {\bibfnamefont {K.}~\bibnamefont {Schwenzer}},\
  }\href {\doibase 10.1103/PhysRevD.81.105010} {\bibfield  {journal} {\bibinfo
  {journal} {Phys. Rev. D}\ }\textbf {\bibinfo {volume} {81}},\ \bibinfo
  {pages} {105010} (\bibinfo {year} {2010})}\BibitemShut {NoStop}%
\bibitem [{\citenamefont {Aguilar}\ \emph {et~al.}(2008)\citenamefont
  {Aguilar}, \citenamefont {Binosi},\ and\ \citenamefont
  {Papavassiliou}}]{Aguilar:2008xm}%
  \BibitemOpen
  \bibfield  {author} {\bibinfo {author} {\bibfnamefont {A.~C.}\ \bibnamefont
  {Aguilar}}, \bibinfo {author} {\bibfnamefont {D.}~\bibnamefont {Binosi}}, \
  and\ \bibinfo {author} {\bibfnamefont {J.}~\bibnamefont {Papavassiliou}},\
  }\href {\doibase 10.1103/PhysRevD.78.025010} {\bibfield  {journal} {\bibinfo
  {journal} {Phys. Rev.}\ }\textbf {\bibinfo {volume} {D78}},\ \bibinfo {pages}
  {025010} (\bibinfo {year} {2008})}\BibitemShut {NoStop}%
\bibitem [{\citenamefont {Vujinovic}\ \emph {et~al.}(2014)\citenamefont
  {Vujinovic}, \citenamefont {Alkofer}, \citenamefont {Eichmann},\ and\
  \citenamefont {Williams}}]{Vujinovic:2014fza}%
  \BibitemOpen
  \bibfield  {author} {\bibinfo {author} {\bibfnamefont {M.}~\bibnamefont
  {Vujinovic}}, \bibinfo {author} {\bibfnamefont {R.}~\bibnamefont {Alkofer}},
  \bibinfo {author} {\bibfnamefont {G.}~\bibnamefont {Eichmann}}, \ and\
  \bibinfo {author} {\bibfnamefont {R.}~\bibnamefont {Williams}},\ }\href
  {\doibase 10.5506/APhysPolBSupp.7.607} {\bibfield  {journal} {\bibinfo
  {journal} {Acta Phys. Polon. Supp.}\ }\textbf {\bibinfo {volume} {7}},\
  \bibinfo {pages} {607} (\bibinfo {year} {2014})}\BibitemShut {NoStop}%
\bibitem [{\citenamefont {Aguilar}\ \emph
  {et~al.}(2021{\natexlab{a}})\citenamefont {Aguilar}, \citenamefont
  {Ferreira},\ and\ \citenamefont {Papavassiliou}}]{Aguilar:2021uwa}%
  \BibitemOpen
  \bibfield  {author} {\bibinfo {author} {\bibfnamefont {A.~C.}\ \bibnamefont
  {Aguilar}}, \bibinfo {author} {\bibfnamefont {M.~N.}\ \bibnamefont
  {Ferreira}}, \ and\ \bibinfo {author} {\bibfnamefont {J.}~\bibnamefont
  {Papavassiliou}},\ }\href{https://arxiv.org/pdf/2111.09431} 
{\ \bibinfo {journal}
 [arXiv:2111.09431 [hep-ph]]}  
 \BibitemShut {NoStop}%
\bibitem [{\citenamefont {Eichmann}\ \emph {et~al.}(2021)\citenamefont
  {Eichmann}, \citenamefont {Pawlowski},\ and\ \citenamefont
  {Silva}}]{Eichmann:2021zuv}%
  \BibitemOpen
  \bibfield  {author} {\bibinfo {author} {\bibfnamefont {G.}~\bibnamefont
  {Eichmann}}, \bibinfo {author} {\bibfnamefont {J.~M.}\ \bibnamefont
  {Pawlowski}}, \ and\ \bibinfo {author} {\bibfnamefont {J.~M.}\
  \bibnamefont {Silva}},\ }\href {\doibase 10.1103/PhysRevD.104.114016}
  {\bibfield  {journal} {\bibinfo  {journal} {Phys. Rev. D}\ }\textbf {\bibinfo
  {volume} {104}},\ \bibinfo {pages} {114016} (\bibinfo {year}
  {2021})}\BibitemShut {NoStop}%
\bibitem [{\citenamefont {Eichmann}\ and\ \citenamefont
  {Pawlowski}(2021)}]{Eichmann:2021miy}%
  \BibitemOpen
  \bibfield  {author} {\bibinfo {author} {\bibfnamefont {G.}~\bibnamefont
  {Eichmann}}\ and\ \bibinfo {author} {\bibfnamefont {J.~M.}\ \bibnamefont
  {Pawlowski}},\ }in\ \href{https://arxiv.org/pdf/2112.08058} 
  {\ \bibinfo {journal}
 [arXiv:2112.08058 [hep-ph]]}
  \BibitemShut {NoStop}%
\bibitem [{\citenamefont {Meyers}\ and\ \citenamefont
  {Swanson}(2013)}]{Meyers:2012ka}%
  \BibitemOpen
  \bibfield  {author} {\bibinfo {author} {\bibfnamefont {J.}~\bibnamefont
  {Meyers}}\ and\ \bibinfo {author} {\bibfnamefont {E.~S.}\ \bibnamefont
  {Swanson}},\ }\href {\doibase 10.1103/PhysRevD.87.036009} {\bibfield
  {journal} {\bibinfo  {journal} {Phys. Rev.}\ }\textbf {\bibinfo {volume}
  {D87}},\ \bibinfo {pages} {036009} (\bibinfo {year} {2013})}\BibitemShut
  {NoStop}%
\bibitem [{\citenamefont {Souza}\ \emph {et~al.}(2020)\citenamefont {Souza},
  \citenamefont {Ferreira}, \citenamefont {Aguilar}, \citenamefont
  {Papavassiliou}, \citenamefont {Roberts},\ and\ \citenamefont
  {Xu}}]{Souza:2019ylx}%
  \BibitemOpen
  \bibfield  {author} {\bibinfo {author} {\bibfnamefont {E.~V.}\ \bibnamefont
  {Souza}}, \bibinfo {author} {\bibfnamefont {M.~N.}\ \bibnamefont {Ferreira}},
  \bibinfo {author} {\bibfnamefont {A.~C.}\ \bibnamefont {Aguilar}}, \bibinfo
  {author} {\bibfnamefont {J.}~\bibnamefont {Papavassiliou}}, \bibinfo {author}
  {\bibfnamefont {C.~D.}\ \bibnamefont {Roberts}}, \ and\ \bibinfo {author}
  {\bibfnamefont {S.-S.}\ \bibnamefont {Xu}},\ }\href {\doibase
  10.1140/epja/s10050-020-00041-y} {\bibfield  {journal} {\bibinfo  {journal}
  {Eur. Phys. J. A}\ }\textbf {\bibinfo {volume} {56}},\ \bibinfo {pages} {25}
  (\bibinfo {year} {2020})}\BibitemShut {NoStop}%
\bibitem [{\citenamefont {Huber}\ \emph {et~al.}(2021)\citenamefont {Huber},
  \citenamefont {Fischer},\ and\ \citenamefont
  {Sanchis-Alepuz}}]{Huber:2021yfy}%
  \BibitemOpen
  \bibfield  {author} {\bibinfo {author} {\bibfnamefont {M.~Q.}\ \bibnamefont
  {Huber}}, \bibinfo {author} {\bibfnamefont {C.~S.}\ \bibnamefont {Fischer}},
  \ and\ \bibinfo {author} {\bibfnamefont {H.}~\bibnamefont {Sanchis-Alepuz}},\
  }\href {\doibase 10.1140/epjc/s10052-021-09864-5} {\bibfield  {journal}
  {\bibinfo  {journal} {Eur. Phys. J. C}\ }\textbf {\bibinfo {volume} {81}},\
  \bibinfo {pages} {1083} (\bibinfo {year} {2021})}\BibitemShut {NoStop}%
\bibitem [{\citenamefont {Aguilar}\ \emph
  {et~al.}(2019{\natexlab{c}})\citenamefont {Aguilar}, \citenamefont
  {Ferreira}, \citenamefont {Figueiredo},\ and\ \citenamefont
  {Papavassiliou}}]{Aguilar:2018csq}%
  \BibitemOpen
  \bibfield  {author} {\bibinfo {author} {\bibfnamefont {A.~C.}\ \bibnamefont
  {Aguilar}}, \bibinfo {author} {\bibfnamefont {M.~N.}\ \bibnamefont
  {Ferreira}}, \bibinfo {author} {\bibfnamefont {C.~T.}\ \bibnamefont
  {Figueiredo}}, \ and\ \bibinfo {author} {\bibfnamefont {J.}~\bibnamefont
  {Papavassiliou}},\ }\href {\doibase 10.1103/PhysRevD.99.034026} {\bibfield
  {journal} {\bibinfo  {journal} {Phys. Rev.}\ }\textbf {\bibinfo {volume}
  {D99}},\ \bibinfo {pages} {034026} (\bibinfo {year}
  {2019}{\natexlab{c}})}\BibitemShut {NoStop}%
\bibitem [{\citenamefont {Schleifenbaum}\ \emph {et~al.}(2005)\citenamefont
  {Schleifenbaum}, \citenamefont {Maas}, \citenamefont {Wambach},\ and\
  \citenamefont {Alkofer}}]{Schleifenbaum:2004id}%
  \BibitemOpen
  \bibfield  {author} {\bibinfo {author} {\bibfnamefont {W.}~\bibnamefont
  {Schleifenbaum}}, \bibinfo {author} {\bibfnamefont {A.}~\bibnamefont {Maas}},
  \bibinfo {author} {\bibfnamefont {J.}~\bibnamefont {Wambach}}, \ and\
  \bibinfo {author} {\bibfnamefont {R.}~\bibnamefont {Alkofer}},\ }\href
  {\doibase 10.1103/PhysRevD.72.014017} {\bibfield  {journal} {\bibinfo
  {journal} {Phys. Rev. D}\ }\textbf {\bibinfo {volume} {72}},\ \bibinfo
  {pages} {014017} (\bibinfo {year} {2005})}\BibitemShut {NoStop}%
\bibitem [{\citenamefont {Huber}\ and\ \citenamefont {von
  Smekal}(2013)}]{Huber:2012kd}%
  \BibitemOpen
  \bibfield  {author} {\bibinfo {author} {\bibfnamefont {M.~Q.}\ \bibnamefont
  {Huber}}\ and\ \bibinfo {author} {\bibfnamefont {L.}~\bibnamefont {von
  Smekal}},\ }\href {\doibase 10.1007/JHEP04(2013)149} {\bibfield  {journal}
  {\bibinfo  {journal} {J. High Energy Phys.}\ }\textbf {\bibinfo {volume}
  {04}},\ \bibinfo {pages} {149} (\bibinfo {year} {2013})}\BibitemShut
  {NoStop}%
\bibitem [{\citenamefont {Aguilar}\ \emph {et~al.}(2013)\citenamefont
  {Aguilar}, \citenamefont {Iba{\~n}ez},\ and\ \citenamefont
  {Papavassiliou}}]{Aguilar:2013xqa}%
  \BibitemOpen
  \bibfield  {author} {\bibinfo {author} {\bibfnamefont {A.~C.}\ \bibnamefont
  {Aguilar}}, \bibinfo {author} {\bibfnamefont {D.}~\bibnamefont {Iba{\~n}ez}},
  \ and\ \bibinfo {author} {\bibfnamefont {J.}~\bibnamefont {Papavassiliou}},\
  }\href {\doibase 10.1103/PhysRevD.87.114020} {\bibfield  {journal} {\bibinfo
  {journal} {Phys. Rev.}\ }\textbf {\bibinfo {volume} {D87}},\ \bibinfo {pages}
  {114020} (\bibinfo {year} {2013})}\BibitemShut {NoStop}%
\bibitem [{\citenamefont {Binosi}\ \emph {et~al.}(2017)\citenamefont {Binosi},
  \citenamefont {Chang}, \citenamefont {Papavassiliou}, \citenamefont {Qin},\
  and\ \citenamefont {Roberts}}]{Binosi:2016wcx}%
  \BibitemOpen
  \bibfield  {author} {\bibinfo {author} {\bibfnamefont {D.}~\bibnamefont
  {Binosi}}, \bibinfo {author} {\bibfnamefont {L.}~\bibnamefont {Chang}},
  \bibinfo {author} {\bibfnamefont {J.}~\bibnamefont {Papavassiliou}}, \bibinfo
  {author} {\bibfnamefont {S.-X.}\ \bibnamefont {Qin}}, \ and\ \bibinfo
  {author} {\bibfnamefont {C.~D.}\ \bibnamefont {Roberts}},\ }\href {\doibase
  10.1103/PhysRevD.95.031501} {\bibfield  {journal} {\bibinfo  {journal} {Phys.
  Rev.}\ }\textbf {\bibinfo {volume} {D95}},\ \bibinfo {pages} {031501}
  (\bibinfo {year} {2017})}\BibitemShut {NoStop}%
\bibitem [{\citenamefont {Cyrol}\ \emph
  {et~al.}(2018{\natexlab{a}})\citenamefont {Cyrol}, \citenamefont {Mitter},
  \citenamefont {Pawlowski},\ and\ \citenamefont {Strodthoff}}]{Cyrol:2017ewj}%
  \BibitemOpen
  \bibfield  {author} {\bibinfo {author} {\bibfnamefont {A.~K.}\ \bibnamefont
  {Cyrol}}, \bibinfo {author} {\bibfnamefont {M.}~\bibnamefont {Mitter}},
  \bibinfo {author} {\bibfnamefont {J.~M.}\ \bibnamefont {Pawlowski}}, \ and\
  \bibinfo {author} {\bibfnamefont {N.}~\bibnamefont {Strodthoff}},\ }\href
  {\doibase 10.1103/PhysRevD.97.054006} {\bibfield  {journal} {\bibinfo
  {journal} {Phys. Rev.}\ }\textbf {\bibinfo {volume} {D97}},\ \bibinfo {pages}
  {054006} (\bibinfo {year} {2018}{\natexlab{a}})}\BibitemShut {NoStop}%
\bibitem [{\citenamefont {Aguilar}\ \emph
  {et~al.}(2021{\natexlab{b}})\citenamefont {Aguilar}, \citenamefont {De~Soto},
  \citenamefont {Ferreira}, \citenamefont {Papavassiliou},\ and\ \citenamefont
  {Rodr\'\i{}guez-Quintero}}]{Aguilar:2021lke}%
  \BibitemOpen
  \bibfield  {author} {\bibinfo {author} {\bibfnamefont {A.~C.}\ \bibnamefont
  {Aguilar}}, \bibinfo {author} {\bibfnamefont {F.}~\bibnamefont {De~Soto}},
  \bibinfo {author} {\bibfnamefont {M.~N.}\ \bibnamefont {Ferreira}}, \bibinfo
  {author} {\bibfnamefont {J.}~\bibnamefont {Papavassiliou}}, \ and\ \bibinfo
  {author} {\bibfnamefont {J.}~\bibnamefont {Rodr\'\i{}guez-Quintero}},\ }\href
  {\doibase 10.1016/j.physletb.2021.136352} {\bibfield  {journal} {\bibinfo
  {journal} {Phys. Lett. B}\ }\textbf {\bibinfo {volume} {818}},\ \bibinfo
  {pages} {136352} (\bibinfo {year} {2021}{\natexlab{b}})}\BibitemShut
  {NoStop}%
\bibitem [{\citenamefont {Salam}(1963)}]{Salam:1963sa}%
  \BibitemOpen
  \bibfield  {author} {\bibinfo {author} {\bibfnamefont {A.}~\bibnamefont
  {Salam}},\ }\href {\doibase 10.1103/PhysRev.130.1287} {\bibfield  {journal}
  {\bibinfo  {journal} {Phys. Rev.}\ }\textbf {\bibinfo {volume} {130}},\
  \bibinfo {pages} {1287} (\bibinfo {year} {1963})}\BibitemShut {NoStop}%
\bibitem [{\citenamefont {Salam}\ and\ \citenamefont
  {Delbourgo}(1964)}]{Salam:1964zk}%
  \BibitemOpen
  \bibfield  {author} {\bibinfo {author} {\bibfnamefont {A.}~\bibnamefont
  {Salam}}\ and\ \bibinfo {author} {\bibfnamefont {R.}~\bibnamefont
  {Delbourgo}},\ }\href {\doibase 10.1103/PhysRev.135.B1398} {\bibfield
  {journal} {\bibinfo  {journal} {Phys. Rev.}\ }\textbf {\bibinfo {volume}
  {135}},\ \bibinfo {pages} {B1398} (\bibinfo {year} {1964})}\BibitemShut
  {NoStop}%
\bibitem [{\citenamefont {Delbourgo}\ and\ \citenamefont
  {West}(1977{\natexlab{a}})}]{Delbourgo:1977jc}%
  \BibitemOpen
  \bibfield  {author} {\bibinfo {author} {\bibfnamefont {R.}~\bibnamefont
  {Delbourgo}}\ and\ \bibinfo {author} {\bibfnamefont {P.~C.}\ \bibnamefont
  {West}},\ }\href {\doibase 10.1088/0305-4470/10/6/024} {\bibfield  {journal}
  {\bibinfo  {journal} {J. Phys. A}\ }\textbf {\bibinfo {volume} {10}},\
  \bibinfo {pages} {1049} (\bibinfo {year} {1977}{\natexlab{a}})}\BibitemShut
  {NoStop}%
\bibitem [{\citenamefont {Delbourgo}\ and\ \citenamefont
  {West}(1977{\natexlab{b}})}]{Delbourgo:1977hq}%
  \BibitemOpen
  \bibfield  {author} {\bibinfo {author} {\bibfnamefont {R.}~\bibnamefont
  {Delbourgo}}\ and\ \bibinfo {author} {\bibfnamefont {P.~C.}\ \bibnamefont
  {West}},\ }\href {\doibase 10.1016/0370-2693(77)90071-5} {\bibfield
  {journal} {\bibinfo  {journal} {Phys. Lett. B}\ }\textbf {\bibinfo {volume}
  {72}},\ \bibinfo {pages} {96} (\bibinfo {year}
  {1977}{\natexlab{b}})}\BibitemShut {NoStop}%
\bibitem [{\citenamefont {Binosi}\ and\ \citenamefont
  {Papavassiliou}(2009)}]{Binosi:2009qm}%
  \BibitemOpen
  \bibfield  {author} {\bibinfo {author} {\bibfnamefont {D.}~\bibnamefont
  {Binosi}}\ and\ \bibinfo {author} {\bibfnamefont {J.}~\bibnamefont
  {Papavassiliou}},\ }\href {\doibase 10.1016/j.physrep.2009.05.001} {\bibfield
   {journal} {\bibinfo  {journal} {Phys. Rept.}\ }\textbf {\bibinfo {volume}
  {479}},\ \bibinfo {pages} {1} (\bibinfo {year} {2009})}\BibitemShut {NoStop}%
\bibitem [{\citenamefont {Roberts}(2020)}]{Roberts:2020hiw}%
  \BibitemOpen
  \bibfield  {author} {\bibinfo {author} {\bibfnamefont {C.~D.}\ \bibnamefont
  {Roberts}},\ }\href {\doibase 10.3390/sym12091468} {\bibfield  {journal}
  {\bibinfo  {journal} {Symmetry}\ }\textbf {\bibinfo {volume} {12}},\ \bibinfo
  {pages} {1468} (\bibinfo {year} {2020})}\BibitemShut {NoStop}%
\bibitem [{\citenamefont {Roberts}\ and\ \citenamefont
  {Williams}(1994)}]{Roberts:1994dr}%
  \BibitemOpen
  \bibfield  {author} {\bibinfo {author} {\bibfnamefont {C.~D.}\ \bibnamefont
  {Roberts}}\ and\ \bibinfo {author} {\bibfnamefont {A.~G.}\ \bibnamefont
  {Williams}},\ }\href {\doibase 10.1016/0146-6410(94)90049-3} {\bibfield
  {journal} {\bibinfo  {journal} {Prog. Part. Nucl. Phys.}\ }\textbf {\bibinfo
  {volume} {33}},\ \bibinfo {pages} {477} (\bibinfo {year} {1994})}\BibitemShut
  {NoStop}%
\bibitem [{\citenamefont {Alkofer}\ and\ \citenamefont {von
  Smekal}(2001)}]{Alkofer:2000wg}%
  \BibitemOpen
  \bibfield  {author} {\bibinfo {author} {\bibfnamefont {R.}~\bibnamefont
  {Alkofer}}\ and\ \bibinfo {author} {\bibfnamefont {L.}~\bibnamefont {von
  Smekal}},\ }\href {\doibase 10.1016/S0370-1573(01)00010-2} {\bibfield
  {journal} {\bibinfo  {journal} {Phys. Rept.}\ }\textbf {\bibinfo {volume}
  {353}},\ \bibinfo {pages} {281} (\bibinfo {year} {2001})}\BibitemShut
  {NoStop}%
\bibitem [{\citenamefont {Fischer}(2006)}]{Fischer:2006ub}%
  \BibitemOpen
  \bibfield  {author} {\bibinfo {author} {\bibfnamefont {C.~S.}\ \bibnamefont
  {Fischer}},\ }\href {\doibase 10.1088/0954-3899/32/8/R02} {\bibfield
  {journal} {\bibinfo  {journal} {J. Phys. G}\ }\textbf {\bibinfo {volume}
  {32}},\ \bibinfo {pages} {R253} (\bibinfo {year} {2006})}\BibitemShut
  {NoStop}%
\bibitem [{\citenamefont {Cloet}\ and\ \citenamefont
  {Roberts}(2014)}]{Cloet:2013jya}%
  \BibitemOpen
  \bibfield  {author} {\bibinfo {author} {\bibfnamefont {I.~C.}\ \bibnamefont
  {Cloet}}\ and\ \bibinfo {author} {\bibfnamefont {C.~D.}\ \bibnamefont
  {Roberts}},\ }\href {\doibase 10.1016/j.ppnp.2014.02.001} {\bibfield
  {journal} {\bibinfo  {journal} {Prog. Part. Nucl. Phys.}\ }\textbf {\bibinfo
  {volume} {77}},\ \bibinfo {pages} {1} (\bibinfo {year} {2014})}\BibitemShut
  {NoStop}%
\bibitem [{\citenamefont {Aguilar}\ \emph {et~al.}(2016)\citenamefont
  {Aguilar}, \citenamefont {Binosi},\ and\ \citenamefont
  {Papavassiliou}}]{Aguilar:2015bud}%
  \BibitemOpen
  \bibfield  {author} {\bibinfo {author} {\bibfnamefont {A.~C.}\ \bibnamefont
  {Aguilar}}, \bibinfo {author} {\bibfnamefont {D.}~\bibnamefont {Binosi}}, \
  and\ \bibinfo {author} {\bibfnamefont {J.}~\bibnamefont {Papavassiliou}},\
  }\href {\doibase 10.1007/s11467-015-0517-6} {\bibfield  {journal} {\bibinfo
  {journal} {Front. Phys.(Beijing)}\ }\textbf {\bibinfo {volume} {11}},\
  \bibinfo {pages} {111203} (\bibinfo {year} {2016})}\BibitemShut {NoStop}%
\bibitem [{\citenamefont {Binosi}\ \emph {et~al.}(2015)\citenamefont {Binosi},
  \citenamefont {Chang}, \citenamefont {Papavassiliou},\ and\ \citenamefont
  {Roberts}}]{Binosi:2014aea}%
  \BibitemOpen
  \bibfield  {author} {\bibinfo {author} {\bibfnamefont {D.}~\bibnamefont
  {Binosi}}, \bibinfo {author} {\bibfnamefont {L.}~\bibnamefont {Chang}},
  \bibinfo {author} {\bibfnamefont {J.}~\bibnamefont {Papavassiliou}}, \ and\
  \bibinfo {author} {\bibfnamefont {C.~D.}\ \bibnamefont {Roberts}},\ }\href
  {\doibase 10.1016/j.physletb.2015.01.031} {\bibfield  {journal} {\bibinfo
  {journal} {Phys. Lett.}\ }\textbf {\bibinfo {volume} {B742}},\ \bibinfo
  {pages} {183} (\bibinfo {year} {2015})}\BibitemShut {NoStop}%
\bibitem [{\citenamefont {Fu}\ \emph {et~al.}(2020)\citenamefont {Fu},
  \citenamefont {Pawlowski},\ and\ \citenamefont {Rennecke}}]{Fu:2019hdw}%
  \BibitemOpen
  \bibfield  {author} {\bibinfo {author} {\bibfnamefont {W.-j.}\ \bibnamefont
  {Fu}}, \bibinfo {author} {\bibfnamefont {J.~M.}\ \bibnamefont {Pawlowski}}, \
  and\ \bibinfo {author} {\bibfnamefont {F.}~\bibnamefont {Rennecke}},\ }\href
  {\doibase 10.1103/PhysRevD.101.054032} {\bibfield  {journal} {\bibinfo
  {journal} {Phys. Rev. D}\ }\textbf {\bibinfo {volume} {101}},\ \bibinfo
  {pages} {054032} (\bibinfo {year} {2020})}\BibitemShut {NoStop}%
\bibitem [{\citenamefont {Aguilar}\ \emph
  {et~al.}(2021{\natexlab{c}})\citenamefont {Aguilar}, \citenamefont
  {Ambr\'osio}, \citenamefont {De~Soto}, \citenamefont {Ferreira},
  \citenamefont {Oliveira}, \citenamefont {Papavassiliou},\ and\ \citenamefont
  {Rodr\'\i{}guez-Quintero}}]{Aguilar:2021okw}%
  \BibitemOpen
  \bibfield  {author} {\bibinfo {author} {\bibfnamefont {A.~C.}\ \bibnamefont
  {Aguilar}}, \bibinfo {author} {\bibfnamefont {C.~O.}\ \bibnamefont
  {Ambr\'osio}}, \bibinfo {author} {\bibfnamefont {F.}~\bibnamefont {De~Soto}},
  \bibinfo {author} {\bibfnamefont {M.~N.}\ \bibnamefont {Ferreira}}, \bibinfo
  {author} {\bibfnamefont {B.~M.}\ \bibnamefont {Oliveira}}, \bibinfo {author}
  {\bibfnamefont {J.}~\bibnamefont {Papavassiliou}}, \ and\ \bibinfo {author}
  {\bibfnamefont {J.}~\bibnamefont {Rodr\'\i{}guez-Quintero}},\ }\href
  {\doibase 10.1103/PhysRevD.104.054028} {\bibfield  {journal} {\bibinfo
  {journal} {Phys. Rev. D}\ }\textbf {\bibinfo {volume} {104}},\ \bibinfo
  {pages} {054028} (\bibinfo {year} {2021}{\natexlab{c}})}\BibitemShut
  {NoStop}%
\bibitem [{\citenamefont {Aguilar}\ \emph {et~al.}(2009)\citenamefont
  {Aguilar}, \citenamefont {Binosi}, \citenamefont {Papavassiliou},\ and\
  \citenamefont {Rodriguez-Quintero}}]{Aguilar:2009nf}%
  \BibitemOpen
  \bibfield  {author} {\bibinfo {author} {\bibfnamefont {A.~C.}\ \bibnamefont
  {Aguilar}}, \bibinfo {author} {\bibfnamefont {D.}~\bibnamefont {Binosi}},
  \bibinfo {author} {\bibfnamefont {J.}~\bibnamefont {Papavassiliou}}, \ and\
  \bibinfo {author} {\bibfnamefont {J.}~\bibnamefont {Rodriguez-Quintero}},\
  }\href {\doibase 10.1103/PhysRevD.80.085018} {\bibfield  {journal} {\bibinfo
  {journal} {Phys. Rev.}\ }\textbf {\bibinfo {volume} {D80}},\ \bibinfo {pages}
  {085018} (\bibinfo {year} {2009})}\BibitemShut {NoStop}%
\bibitem [{\citenamefont {Boucaud}\ \emph {et~al.}(2009)\citenamefont
  {Boucaud}, \citenamefont {De~Soto}, \citenamefont {Leroy}, \citenamefont
  {Le~Yaouanc}, \citenamefont {Micheli} \emph {et~al.}}]{Boucaud:2008gn}%
  \BibitemOpen
  \bibfield  {author} {\bibinfo {author} {\bibfnamefont {P.}~\bibnamefont
  {Boucaud}}, \bibinfo {author} {\bibfnamefont {F.}~\bibnamefont {De~Soto}},
  \bibinfo {author} {\bibfnamefont {J.}~\bibnamefont {Leroy}}, \bibinfo
  {author} {\bibfnamefont {A.}~\bibnamefont {Le~Yaouanc}}, \bibinfo {author}
  {\bibfnamefont {J.}~\bibnamefont {Micheli}},  \emph {et~al.},\ }\href
  {\doibase 10.1103/PhysRevD.79.014508} {\bibfield  {journal} {\bibinfo
  {journal} {Phys. Rev.}\ }\textbf {\bibinfo {volume} {D79}},\ \bibinfo {pages}
  {014508} (\bibinfo {year} {2009})}\BibitemShut {NoStop}%
\bibitem [{\citenamefont {Boucaud}\ \emph {et~al.}(2011)\citenamefont
  {Boucaud}, \citenamefont {Dudal}, \citenamefont {Leroy}, \citenamefont
  {Pene},\ and\ \citenamefont {Rodriguez-Quintero}}]{Boucaud:2011eh}%
  \BibitemOpen
  \bibfield  {author} {\bibinfo {author} {\bibfnamefont {P.}~\bibnamefont
  {Boucaud}}, \bibinfo {author} {\bibfnamefont {D.}~\bibnamefont {Dudal}},
  \bibinfo {author} {\bibfnamefont {J.}~\bibnamefont {Leroy}}, \bibinfo
  {author} {\bibfnamefont {O.}~\bibnamefont {Pene}}, \ and\ \bibinfo {author}
  {\bibfnamefont {J.}~\bibnamefont {Rodriguez-Quintero}},\ }\href {\doibase
  10.1007/JHEP12(2011)018} {\bibfield  {journal} {\bibinfo  {journal} {J. High
  Energy Phys.}\ }\textbf {\bibinfo {volume} {12}},\ \bibinfo {pages} {018}
  (\bibinfo {year} {2011})}\BibitemShut {NoStop}%
\bibitem [{\citenamefont {von Smekal}\ \emph {et~al.}(2009)\citenamefont {von
  Smekal}, \citenamefont {Maltman},\ and\ \citenamefont
  {Sternbeck}}]{vonSmekal:2009ae}%
  \BibitemOpen
  \bibfield  {author} {\bibinfo {author} {\bibfnamefont {L.}~\bibnamefont {von
  Smekal}}, \bibinfo {author} {\bibfnamefont {K.}~\bibnamefont {Maltman}}, \
  and\ \bibinfo {author} {\bibfnamefont {A.}~\bibnamefont {Sternbeck}},\ }\href
  {\doibase 10.1016/j.physletb.2009.10.030} {\bibfield  {journal} {\bibinfo
  {journal} {Phys. Lett. B}\ }\textbf {\bibinfo {volume} {681}},\ \bibinfo
  {pages} {336} (\bibinfo {year} {2009})}\BibitemShut {NoStop}%
\bibitem [{\citenamefont {Williams}(2015)}]{Williams:2014iea}%
  \BibitemOpen
  \bibfield  {author} {\bibinfo {author} {\bibfnamefont {R.}~\bibnamefont
  {Williams}},\ }\href {\doibase 10.1140/epja/i2015-15057-4} {\bibfield
  {journal} {\bibinfo  {journal} {Eur. Phys. J.}\ }\textbf {\bibinfo {volume}
  {A51}},\ \bibinfo {pages} {57} (\bibinfo {year} {2015})}\BibitemShut
  {NoStop}%
\bibitem [{\citenamefont {Aguilar}\ \emph
  {et~al.}(2020{\natexlab{b}})\citenamefont {Aguilar}, \citenamefont
  {Ferreira},\ and\ \citenamefont {Papavassiliou}}]{Aguilar:2020yni}%
  \BibitemOpen
  \bibfield  {author} {\bibinfo {author} {\bibfnamefont {A.~C.}\ \bibnamefont
  {Aguilar}}, \bibinfo {author} {\bibfnamefont {M.~N.}\ \bibnamefont
  {Ferreira}}, \ and\ \bibinfo {author} {\bibfnamefont {J.}~\bibnamefont
  {Papavassiliou}},\ }\href {\doibase 10.1140/epjc/s10052-020-08453-2}
  {\bibfield  {journal} {\bibinfo  {journal} {Eur. Phys. J. C}\ }\textbf
  {\bibinfo {volume} {80}},\ \bibinfo {pages} {887} (\bibinfo {year}
  {2020}{\natexlab{b}})}\BibitemShut {NoStop}%
\bibitem [{\citenamefont {Morningstar}\ and\ \citenamefont
  {Peardon}(1999)}]{Morningstar:1999rf}%
  \BibitemOpen
  \bibfield  {author} {\bibinfo {author} {\bibfnamefont {C.~J.}\ \bibnamefont
  {Morningstar}}\ and\ \bibinfo {author} {\bibfnamefont {M.~J.}\ \bibnamefont
  {Peardon}},\ }\href {\doibase 10.1103/PhysRevD.60.034509} {\bibfield
  {journal} {\bibinfo  {journal} {Phys. Rev.}\ }\textbf {\bibinfo {volume}
  {D60}},\ \bibinfo {pages} {034509} (\bibinfo {year} {1999})}\BibitemShut
  {NoStop}%
\bibitem [{\citenamefont {Bali}\ \emph {et~al.}(1993)\citenamefont {Bali},
  \citenamefont {Schilling}, \citenamefont {Hulsebos}, \citenamefont {Irving},
  \citenamefont {Michael},\ and\ \citenamefont {Stephenson}}]{Bali:1993fb}%
  \BibitemOpen
  \bibfield  {author} {\bibinfo {author} {\bibfnamefont {G.~S.}\ \bibnamefont
  {Bali}}, \bibinfo {author} {\bibfnamefont {K.}~\bibnamefont {Schilling}},
  \bibinfo {author} {\bibfnamefont {A.}~\bibnamefont {Hulsebos}}, \bibinfo
  {author} {\bibfnamefont {A.~C.}\ \bibnamefont {Irving}}, \bibinfo {author}
  {\bibfnamefont {C.}~\bibnamefont {Michael}}, \ and\ \bibinfo {author}
  {\bibfnamefont {P.~W.}\ \bibnamefont {Stephenson}} (\bibinfo {collaboration}
  {UKQCD}),\ }\href {\doibase 10.1016/0370-2693(93)90948-H} {\bibfield
  {journal} {\bibinfo  {journal} {Phys. Lett. B}\ }\textbf {\bibinfo {volume}
  {309}},\ \bibinfo {pages} {378} (\bibinfo {year} {1993})}\BibitemShut
  {NoStop}%
\bibitem [{\citenamefont {McNeile}(2009)}]{McNeile:2008sr}%
  \BibitemOpen
  \bibfield  {author} {\bibinfo {author} {\bibfnamefont {C.}~\bibnamefont
  {McNeile}},\ }\href {\doibase 10.1016/j.nuclphysbps.2008.12.059} {\bibfield
  {journal} {\bibinfo  {journal} {Nucl. Phys. B Proc. Suppl.}\ }\textbf
  {\bibinfo {volume} {186}},\ \bibinfo {pages} {264} (\bibinfo {year}
  {2009})}\BibitemShut {NoStop}%
\bibitem [{\citenamefont {Chen}\ \emph {et~al.}(2006)\citenamefont {Chen} \emph
  {et~al.}}]{Chen:2005mg}%
  \BibitemOpen
  \bibfield  {author} {\bibinfo {author} {\bibfnamefont {Y.}~\bibnamefont
  {Chen}} \emph {et~al.},\ }\href {\doibase 10.1103/PhysRevD.73.014516}
  {\bibfield  {journal} {\bibinfo  {journal} {Phys. Rev. D}\ }\textbf {\bibinfo
  {volume} {73}},\ \bibinfo {pages} {014516} (\bibinfo {year}
  {2006})}\BibitemShut {NoStop}%
\bibitem [{\citenamefont {Athenodorou}\ and\ \citenamefont
  {Teper}(2020)}]{Athenodorou:2020ani}%
  \BibitemOpen
  \bibfield  {author} {\bibinfo {author} {\bibfnamefont {A.}~\bibnamefont
  {Athenodorou}}\ and\ \bibinfo {author} {\bibfnamefont {M.}~\bibnamefont
  {Teper}},\ }\href {\doibase 10.1007/JHEP11(2020)172} {\bibfield  {journal}
  {\bibinfo  {journal} {JHEP}\ }\textbf {\bibinfo {volume} {11}},\ \bibinfo
  {pages} {172} (\bibinfo {year} {2020})}\BibitemShut {NoStop}%
\bibitem [{\citenamefont {Dudal}\ \emph {et~al.}(2011)\citenamefont {Dudal},
  \citenamefont {Guimaraes},\ and\ \citenamefont {Sorella}}]{Dudal:2010cd}%
  \BibitemOpen
  \bibfield  {author} {\bibinfo {author} {\bibfnamefont {D.}~\bibnamefont
  {Dudal}}, \bibinfo {author} {\bibfnamefont {M.~S.}\ \bibnamefont
  {Guimaraes}}, \ and\ \bibinfo {author} {\bibfnamefont {S.~P.}\ \bibnamefont
  {Sorella}},\ }\href {\doibase 10.1103/PhysRevLett.106.062003} {\bibfield
  {journal} {\bibinfo  {journal} {Phys. Rev. Lett.}\ }\textbf {\bibinfo
  {volume} {106}},\ \bibinfo {pages} {062003} (\bibinfo {year}
  {2011})}\BibitemShut {NoStop}%
\bibitem [{\citenamefont {Xu}\ \emph {et~al.}(2019)\citenamefont {Xu},
  \citenamefont {Cui}, \citenamefont {Chang}, \citenamefont {Papavassiliou},
  \citenamefont {Roberts},\ and\ \citenamefont {Zong}}]{Xu:2018cor}%
  \BibitemOpen
  \bibfield  {author} {\bibinfo {author} {\bibfnamefont {S.-S.}\ \bibnamefont
  {Xu}}, \bibinfo {author} {\bibfnamefont {Z.-F.}\ \bibnamefont {Cui}},
  \bibinfo {author} {\bibfnamefont {L.}~\bibnamefont {Chang}}, \bibinfo
  {author} {\bibfnamefont {J.}~\bibnamefont {Papavassiliou}}, \bibinfo {author}
  {\bibfnamefont {C.~D.}\ \bibnamefont {Roberts}}, \ and\ \bibinfo {author}
  {\bibfnamefont {H.-S.}\ \bibnamefont {Zong}},\ }\href {\doibase
  10.1140/epja/i2019-12805-4} {\bibfield  {journal} {\bibinfo  {journal} {Eur.
  Phys. J.}\ }\textbf {\bibinfo {volume} {A55}},\ \bibinfo {pages} {113}
  (\bibinfo {year} {2019})}\BibitemShut {NoStop}%
\bibitem [{\citenamefont {Schwinger}(1962{\natexlab{a}})}]{Schwinger:1962tn}%
  \BibitemOpen
  \bibfield  {author} {\bibinfo {author} {\bibfnamefont {J.~S.}\ \bibnamefont
  {Schwinger}},\ }\href {\doibase 10.1103/PhysRev.125.397} {\bibfield
  {journal} {\bibinfo  {journal} {Phys. Rev.}\ }\textbf {\bibinfo {volume}
  {125}},\ \bibinfo {pages} {397} (\bibinfo {year}
  {1962}{\natexlab{a}})}\BibitemShut {NoStop}%
\bibitem [{\citenamefont {Schwinger}(1962{\natexlab{b}})}]{Schwinger:1962tp}%
  \BibitemOpen
  \bibfield  {author} {\bibinfo {author} {\bibfnamefont {J.~S.}\ \bibnamefont
  {Schwinger}},\ }\href {\doibase 10.1103/PhysRev.128.2425} {\bibfield
  {journal} {\bibinfo  {journal} {Phys. Rev.}\ }\textbf {\bibinfo {volume}
  {128}},\ \bibinfo {pages} {2425} (\bibinfo {year}
  {1962}{\natexlab{b}})}\BibitemShut {NoStop}%
\bibitem [{\citenamefont {Aguilar}\ \emph {et~al.}(2012)\citenamefont
  {Aguilar}, \citenamefont {Ibanez}, \citenamefont {Mathieu},\ and\
  \citenamefont {Papavassiliou}}]{Aguilar:2011xe}%
  \BibitemOpen
  \bibfield  {author} {\bibinfo {author} {\bibfnamefont {A.~C.}\ \bibnamefont
  {Aguilar}}, \bibinfo {author} {\bibfnamefont {D.}~\bibnamefont {Ibanez}},
  \bibinfo {author} {\bibfnamefont {V.}~\bibnamefont {Mathieu}}, \ and\
  \bibinfo {author} {\bibfnamefont {J.}~\bibnamefont {Papavassiliou}},\ }\href
  {\doibase 10.1103/PhysRevD.85.014018} {\bibfield  {journal} {\bibinfo
  {journal} {Phys. Rev.}\ }\textbf {\bibinfo {volume} {D85}},\ \bibinfo {pages}
  {014018} (\bibinfo {year} {2012})}\BibitemShut {NoStop}%
\bibitem [{\citenamefont {Iba{\~n}ez}\ and\ \citenamefont
  {Papavassiliou}(2013)}]{Ibanez:2012zk}%
  \BibitemOpen
  \bibfield  {author} {\bibinfo {author} {\bibfnamefont {D.}~\bibnamefont
  {Iba{\~n}ez}}\ and\ \bibinfo {author} {\bibfnamefont {J.}~\bibnamefont
  {Papavassiliou}},\ }\href {\doibase 10.1103/PhysRevD.87.034008} {\bibfield
  {journal} {\bibinfo  {journal} {Phys. Rev.}\ }\textbf {\bibinfo {volume}
  {D87}},\ \bibinfo {pages} {034008} (\bibinfo {year} {2013})}\BibitemShut
  {NoStop}%
\bibitem [{\citenamefont {Aguilar}\ \emph {et~al.}(2018)\citenamefont
  {Aguilar}, \citenamefont {Binosi}, \citenamefont {Figueiredo},\ and\
  \citenamefont {Papavassiliou}}]{Aguilar:2017dco}%
  \BibitemOpen
  \bibfield  {author} {\bibinfo {author} {\bibfnamefont {A.~C.}\ \bibnamefont
  {Aguilar}}, \bibinfo {author} {\bibfnamefont {D.}~\bibnamefont {Binosi}},
  \bibinfo {author} {\bibfnamefont {C.~T.}\ \bibnamefont {Figueiredo}}, \ and\
  \bibinfo {author} {\bibfnamefont {J.}~\bibnamefont {Papavassiliou}},\ }\href
  {\doibase 10.1140/epjc/s10052-018-5679-2} {\bibfield  {journal} {\bibinfo
  {journal} {Eur. Phys. J.}\ }\textbf {\bibinfo {volume} {C78}},\ \bibinfo
  {pages} {181} (\bibinfo {year} {2018})}\BibitemShut {NoStop}%
\bibitem [{\citenamefont {Binosi}\ and\ \citenamefont
  {Papavassiliou}(2018)}]{Binosi:2017rwj}%
  \BibitemOpen
  \bibfield  {author} {\bibinfo {author} {\bibfnamefont {D.}~\bibnamefont
  {Binosi}}\ and\ \bibinfo {author} {\bibfnamefont {J.}~\bibnamefont
  {Papavassiliou}},\ }\href {\doibase 10.1103/PhysRevD.97.054029} {\bibfield
  {journal} {\bibinfo  {journal} {Phys. Rev.}\ }\textbf {\bibinfo {volume}
  {D97}},\ \bibinfo {pages} {054029} (\bibinfo {year} {2018})}\BibitemShut
  {NoStop}%
\bibitem [{\citenamefont {Cyrol}\ \emph
  {et~al.}(2018{\natexlab{b}})\citenamefont {Cyrol}, \citenamefont {Pawlowski},
  \citenamefont {Rothkopf},\ and\ \citenamefont {Wink}}]{Cyrol:2018xeq}%
  \BibitemOpen
  \bibfield  {author} {\bibinfo {author} {\bibfnamefont {A.~K.}\ \bibnamefont
  {Cyrol}}, \bibinfo {author} {\bibfnamefont {J.~M.}\ \bibnamefont
  {Pawlowski}}, \bibinfo {author} {\bibfnamefont {A.}~\bibnamefont {Rothkopf}},
  \ and\ \bibinfo {author} {\bibfnamefont {N.}~\bibnamefont {Wink}},\ }\href
  {\doibase 10.21468/SciPostPhys.5.6.065} {\bibfield  {journal} {\bibinfo
  {journal} {SciPost Phys.}\ }\textbf {\bibinfo {volume} {5}},\ \bibinfo
  {pages} {065} (\bibinfo {year} {2018}{\natexlab{b}})}\BibitemShut {NoStop}%
\bibitem [{\citenamefont {Horak}\ \emph {et~al.}(2020)\citenamefont {Horak},
  \citenamefont {Pawlowski},\ and\ \citenamefont {Wink}}]{Horak:2020eng}%
  \BibitemOpen
  \bibfield  {author} {\bibinfo {author} {\bibfnamefont {J.}~\bibnamefont
  {Horak}}, \bibinfo {author} {\bibfnamefont {J.~M.}\ \bibnamefont
  {Pawlowski}}, \ and\ \bibinfo {author} {\bibfnamefont {N.}~\bibnamefont
  {Wink}},\ }\href {\doibase 10.1103/PhysRevD.102.125016} {\bibfield  {journal}
  {\bibinfo  {journal} {Phys. Rev. D}\ }\textbf {\bibinfo {volume} {102}},\
  \bibinfo {pages} {125016} (\bibinfo {year} {2020})}\BibitemShut {NoStop}%
\bibitem [{\citenamefont {Horak}\ \emph
  {et~al.}(2021{\natexlab{a}})\citenamefont {Horak}, \citenamefont
  {Papavassiliou}, \citenamefont {Pawlowski},\ and\ \citenamefont
  {Wink}}]{Horak:2021pfr}%
  \BibitemOpen
  \bibfield  {author} {\bibinfo {author} {\bibfnamefont {J.}~\bibnamefont
  {Horak}}, \bibinfo {author} {\bibfnamefont {J.}~\bibnamefont
  {Papavassiliou}}, \bibinfo {author} {\bibfnamefont {J.~M.}\ \bibnamefont
  {Pawlowski}}, \ and\ \bibinfo {author} {\bibfnamefont {N.}~\bibnamefont
  {Wink}},\ }\href {\doibase 10.1103/PhysRevD.104.074017} {\bibfield  {journal}
  {\bibinfo  {journal} {Phys. Rev. D}\ }\textbf {\bibinfo {volume} {104}},\
  \bibinfo {pages} {074017} (\bibinfo {year} {2021}{\natexlab{a}})}\BibitemShut
  {NoStop}%
\bibitem [{\citenamefont {Horak}\ \emph
  {et~al.}(2021{\natexlab{b}})\citenamefont {Horak}, \citenamefont {Pawlowski},
  \citenamefont {Rodr\'\i{}guez-Quintero}, \citenamefont {Turnwald},
  \citenamefont {Urban}, \citenamefont {Wink},\ and\ \citenamefont
  {Zafeiropoulos}}]{Horak:2021syv}%
  \BibitemOpen
  \bibfield  {author} {\bibinfo {author} {\bibfnamefont {J.}~\bibnamefont
  {Horak}}, \bibinfo {author} {\bibfnamefont {J.~M.}\ \bibnamefont
  {Pawlowski}}, \bibinfo {author} {\bibfnamefont {J.}~\bibnamefont
  {Rodr\'\i{}guez-Quintero}}, \bibinfo {author} {\bibfnamefont
  {J.}~\bibnamefont {Turnwald}}, \bibinfo {author} {\bibfnamefont {J.~M.}\
  \bibnamefont {Urban}}, \bibinfo {author} {\bibfnamefont {N.}~\bibnamefont
  {Wink}}, \ and\ \bibinfo {author} {\bibfnamefont {S.}~\bibnamefont
  {Zafeiropoulos}},\ }\href {https://arxiv.org/abs/2107.13464}
   {\ \bibinfo {journal}
  {arXiv:2107.13464 [hep-ph]].}{\natexlab{b}}}\BibitemShut {NoStop}%

\end{thebibliography}

%

\end{document}